\documentclass{article}

\usepackage{PRIMEarxiv}

\usepackage[utf8]{inputenc} % allow utf-8 input
\usepackage[T1]{fontenc}    % use 8-bit T1 fonts
\usepackage{hyperref}       % hyperlinks
\usepackage{url}            % simple URL typesetting
\usepackage{booktabs}       % professional-quality tables
\usepackage{amsfonts}       % blackboard math symbols
\usepackage{nicefrac}       % compact symbols for 1/2, etc.
\usepackage{microtype}      % microtypography
\usepackage{lipsum}
\usepackage{fancyhdr}       % header
\usepackage{graphicx}       % graphics
\graphicspath{{media/}}     % organize your images and other figures under media/ folder

\usepackage{rotating}
\usepackage{ifpdf}
\usepackage[T1]{fontenc}
\usepackage{times}
\usepackage{sourcesanspro}
\usepackage{newtxmath}
\usepackage{textcomp}%
\usepackage{xcolor}%
\usepackage{hyperref}
\usepackage{appendix}
\usepackage{url}
\usepackage{xcolor}
\usepackage{tikz}

\DeclareRobustCommand\dotted{\tikz[baseline=-0.6ex]\draw[thick,dotted] (0,0)--(0.54,0);}
\DeclareRobustCommand\dashed{\tikz[baseline=-0.6ex]\draw[thick,dashed] (0,0)--(0.54,0);}

\usepackage[outdir=./]{epstopdf}
\title{Performance and accuracy assessments of an incompressible fluid solver coupled with a deep convolutional neural network }

\author{
  Ekhi Ajuria Illarramendi\\
  ISAE-SUPAERO / CERFACS \\
  Université de Toulouse\\
  Toulouse (France)\\
  \texttt{ajuria@cerfacs.fr} \\
   \And
  Michaël Bauerheim\\
  ISAE-SUPAERO \\
  Université de Toulouse\\
  Toulouse (France)\\
  \texttt{michael.bauerheim@isae-supaero.fr} \\
  \And
  Bénédicte Cuenot \\
  CERFACS \\
  Toulouse (France)\\
  \texttt{cuenot@cerfacs.fr} \\
}

\begin{document}
\maketitle

\begin{abstract}
The resolution of the Poisson equation is usually one of the most computationally intensive steps for incompressible fluid solvers. Lately, Deep Learning, and especially Convolutional Neural Networks (CNN), has been introduced to solve this equation, leading to significant inference time reduction at the cost of a lack of guarantee on the accuracy of the solution. This drawback might lead to inaccuracies and potentially unstable simulations. It also makes impossible a fair assessment of the CNN speedup, for instance, when changing the network architecture, since evaluated at different error levels. To circumvent this issue, a hybrid strategy is developed, which couples a CNN with a traditional iterative solver to ensure a user-defined accuracy level. The CNN hybrid method  is tested on  two  flow cases, consisting of a variable-density plume with and without obstacles, demostrating remarkable generalization capabilities, ensuring both the accuracy and stability of the simulations. The error distribution of the predictions using several network architectures is further investigated. Results show that the threshold of the hybrid strategy defined as the mean divergence of the velocity field is ensuring a consistent physical behavior of the CNN-based hybrid computational strategy. This strategy allows a systematic evaluation of the CNN performance at the same accuracy level for various network architectures. In particular, the importance of incorporating multiple scales in the network architecture is demonstrated, since improving both the accuracy and the inference performance compared with feedforward CNN architectures, as these networks can provide solutions 10-25 faster than traditional iterative solvers.\footnote{The code used to replicate the results is publically available at \url{https://gitlab.isae-supaero.fr/daep/fluidnet_supaero.git}}.
\end{abstract}

% keywords can be removed

\section{Introduction}

A wide variety of problems encountered in physics, engineering, and medicine among others, can be described with Partial Differential Equations (PDEs). These equations correspond to the mathematical translation of physical laws established from observable data. Among many examples, fluid flows can be described with the Navier-Stokes equations, electric and magnetic fields are modeled with the unified Maxwell's equations, and the variation of concentration of chemicals can be modeled using the reaction-diffusion systems. Except for simplified configurations, PDEs cannot be solved analytically, and require numerical tools to approximate their solutions, which may result in computationally expensive calculations. Focusing on fluid mechanics, the Navier-Stokes equations are a non-linear coupled PDE system which represents flows with a wide range of spatial and temporal scales, governed by forces depending on dimensionless numbers such as the Reynolds, Froude, Mach and Richardson numbers among many. The main difficulty arises from the non-linear term of the momentum equation, which involves spatial gradients. Whereas this term generates multiple complex behaviors of the flow field (vorticity generation, transition to turbulence etc.), it is by definition highly sensitive to the numerical setup: a small error on the gradient estimation may lead to very large differences in the solution. As a result, the CFD community has produced an extreme effort over the last decades to improve numerical schemes~\cite{deng2012} towards high-order discretization with high accuracy, most often at a price of a very high CPU cost. Such simulations have been made possible by massively parallel computing (High Performance Computing, HPC) combined with a constant amelioration of the computational hardware. More recently, the development and improvement of Graphical Processing Units (GPUs) has also enabled to accelerate calculation codes, and has opened the path to Machine Learning (ML) techniques that were not yet developed in the CFD community. 

The theoretical foundations of ML models were developed in the mid-XX-th century. In 1958 Rosenblatt~\cite{rosenblatt1958perceptron}  developed an Artificial Neural Network (\textit{Perceptron}), while the \textit{backpropagation} was introduced in 1960 by Kelley~\cite{kelley1960} for control applications (although the first successfully application was obtained by Rumelhart et al.~\cite{rumelhart1986learning} in 1986). Convolutional Neural Networks (CNNs) appeared slightly later, when LeCun et al.~\cite{lecun1998gradient} created a network that extracted spatial features from 2D images.  The performance of CNNs on image classification problems~\cite{AlexNet2012} and the development of more efficient GPUs then sky-rocketed the use of ML in various domains. Mostly driven by the computer vision community, open-source frameworks were developed, which led to an even further spread of Neural Networks (NNs). This recent \textit{data-driven} wave finally touched the CFD community, who made substantial efforts in the past 5 years to apply ML to flow problems. 

ML techniques are optimization algorithms that extract patterns and structures from data, "generalizing" the valuable information found on its training data. When applied to Fluid dynamics, this usually translates into the creation of Reduced-order Models (ROM), a simple representation of the information contained in the dataset. These simpler representations can be used in two different ways in CFD, as reviewed by Brunton et al.~\cite{brunton2019}: either discovering unknown models, or improving existing models. Discovering unknown models from data has been a long-standing problem in the scientific community. Throughout the history of humanity, people built models to describe the behavior of observed phenomena, in order to make future predictions. Nowadays, with the advance of technology, a time has reached where a lot of data of different processes are available, yet being still under-exploited. This has promoted Machine Learning, with the aim to benefit from these data, by extracting features to develop reduced models. For instance, Brunton et al.~\cite{brunton2016discovering}, followed by Rudy et al.~\cite{rudy2017data}, established governing equations for the chaotic Lorenz system or for the vortex shedding behind a cylinder using a sparse dynamics model and a sparse regression model. Another example is the use of Machine Learning for problems where the governing equations are well known, but too expensive to resolve, as in turbulent flows, for example Zhang et al.~\cite{zhang2015machine} used neural networks to predict subgrid-scale correction factors on turbulent flows. Fukami et al.~\cite{fukami2019} applied CNNs to reconstruct turbulent flows that respect the energy spectrum, while other works~\cite{Xie2018,Kim2019,subramaniam2020turbulence} used Generative Adversarial Networks (GANs) to reconstruct turbulent fields. Similarly, CNNs have also been used in turbulent combustion to predict the sub-grid flame wrinkling~\cite{lapeyre2019training}. 

When the resolution of PDEs is computationally expensive, Machine Learning techniques can be used to develop surrogate models that capture the essence of the studied phenomena with a reduced computational cost. Classical reduced-order models,~\cite{cizmas2008acceleration,rowley2017model,taira2017modal} include proper orthogonal decomposition (POD) or dynamic mode decomposition (DMD). These methods can be combined with Machine Learning techniques, such as autoencoders~\cite{otto2019linearly} or long short-term memory (LSTM) neural networks~\cite{pawar2019data} in order to get satisfactory results on problems like the 2D Navier-Stokes equations. The first studies introducing ML to solve PDEs date back to 1990, with the work of Lee et al.~\cite{lee1990neural} followed by Dissanayake et al.~\cite{Dissanayake1994} in 1994. Dissanayake's team proposed a multilayer feedforward network to solve a linear Poisson equation and a thermal conduction problem on a non-linear heat generation test case. Neural Networks can be described as `universal approximator' functions~\cite{hornik1989}, which enable to easily differentiate their inputs with respect to their output. That way, they constituted a novel approach where the classical finite-element or finite-difference discretization problem was substituted by an unconstrained optimization problem. Following the reduced-order-modeling philosophy, their objective was to obtain an easy-to-implement solver that could rapidly give accurate results. Lagaris et al.~\cite{Lagaris1998} further exploited the idea, adding both Neumann and Dirichlet boundary conditions. Gonzalez-Garcia et al.~\cite{Gonzalez1998} developed a different approach, where they modeled the right-hand side of a PDE with a neural network, in order to turn the problem into a more simple Ordinary Differential Equation. They showed that the network enables to generalize from experimental data which might be incomplete, or noisy, testing on the Kuramoto-Shivanisky equation. All these works profit from the differentiable nature of neural networks. However, these networks were rapidly limited by the available computational capability and were restricted to a single particular PDE, with not much margin for generalization.

With the increase of computational resources, later work with larger and more complex networks~\cite{shirvany2009multilayer}, focused on Multi-Layer Perceptrons (MLP)~\cite{shirvany2009multilayer} or Radial Basis Functions (RBF)~\cite{mai2005solving}. Recently, Raissi et al.~\cite{Raissi2017} introduced \textit{physics informed neural networks} (PINN), where a-priori physical knowledge is introduced into the training process of neural networks. By using the residual of the physical equation to optimize the network's trainable parameters, the network ensures physical laws and may even lead to training processes which do not need a database~\cite{sun2020}. Such approach was tested on different PDEs, including the incompressible Navier Stokes-equations~\cite{raissi2017_2,Raissi2018}, which have also been tackled by several works which employ networks to predict the future state of fluid flows, i.e. as end-to-end solvers. Some recent works include recurrent network architectures, where a LSTM (Long Short-Term Memory~\cite{hochreiter1997long}) network was used to encode fluid structures into a latent space~\cite{wiewel2019latent}, or more classical CNN, where an unsupervised network training is evaluated to test the performance gain of CNNs~\cite{wandel2020unsupervised}.  However, the majority of previously described works, that focused both on the Navier-Stokes or more general PDEs, handle the resolution of PDEs independently to traditional fluid solvers. Even if physical laws are embedded during the training procedure, the interaction between the Neural Networks and traditional fluid solvers still offers a wide variety of options for further studies. One of the study fields on which a large effort is devoted to further study the coupling networks with fluid solvers corresponds to the resolution of the Poisson in incompressible solvers. The problem was first tackled by Yang et al.~\cite{Yang2016}, who used a Multi-Layer Perceptron, and Xiao et al.~\cite{Xiao2018} continued with the idea using CNN in larger domains.  Along the same line, the work of Tompson et al.~\cite{Tompson2017} should be highlighted, where a CNN is coupled to an incompressible fluid solver. They introduced a physical loss similar to the one of PINN, based on the continuity equation. Finally, recent work by Um et al.~\cite{um2020solver} coupled a differentiable fluid solver with a CNN in order to further encode flow dynamics during the network training. 

This work further extends a previous work~\cite{ajuria2020towards}, which introduced a hybrid CNN-Jacobi method to solve incompressible flows, that guarantees a user-defined divergence level. The numerical setup and test cases are described in Section~\ref{sec:plume_cases}, and the CNN approach to solve the Poisson equation is described in Section~\ref{sec:DL_poisson}. Different neural networks are compared for the plume test case, and their behavior and error distribution are compared and analysed in Section~\ref{sec:errorquant}. Finally, Section~\ref{sec:performance} presents a performance assessment of the hybrid CNN approach with various network architectures. Results show that for the same error level, networks with multiple scales allow to perform accurate predictions faster than classical solvers.

\section{Configuration and methodology}

\subsection{Plume Case}
\label{sec:plume_cases}

%{\color{red} c'est trop rapide, met plus de contexte. Commence a dire pourquoi le buyoanci flow sont interessants en meca flu et dans la vraie vie en general. Ensuite parle des plume comme cas simplifie permettant des becnhmark et une comprehension de la physique complexe, couplant grandient de densite et ecoulement tourbillonaire ... et pour tout ca il faut citer des papiers et etudes. Fini a la toute fin en parlant du fait que le ML a attaquer un peu ce probleme pour aller vite pour les simu de fumee par exemple, mais ca doit apparaitre a la fin uniquement}

Buoyancy forces are induced by density variations in the presence of gravity. The study of buoyancy-driven flows has been a long-standing problem for the fluid mechanics community~\cite{turner1979buoyancy}, as they correspond to a wide variety of remarkable geophysical flows, from volcanic eruptions to avalanches. The Earth's climate and ocean's behavior are strongly dependent of buoyancy, as the ocean's currents and tides depend on the creation and distribution of water density variations. Buoyant coastal currents for example, are responsible of redistributing the fresh water coming from rivers and other sources into the ocean, carrying as well sediments or pollutants which can strongly impact natural ecosystems. A thorough analysis of buoyancy-driven flows, especially focused on oceanic flows, can be found in the work of~Chassignet et al.~\cite{chassignet2012buoyancy}. The civil engineering community also studies flows driven by buoyancy forces, which control the ventilation of buildings~\cite{allocca2003design} or even entire cities, with direct effect on the propagation of pollutant particles~\cite{luo2011passive}. The propagation of fires in buildings has also led to extended work with theoretical, numerical and experimental models for smoke propagation, as shown by Ahn~et~al.~\cite{ahn2019theoretical}. 

Usually the term buoyancy-driven flow refers to the situation where a lighter moving fluid is surrounded by a heavier fluid, whereas a heavier moving fluid surrounded by a lighter fluid is called a density-driven flow. This work focuses on plumes, which result from injecting a light fluid into a heavier quiescent environment. Plumes can be found in a wide variety of physical phenomena, such as the flow escaping the chimney of a locomotive steam engine or even the smoke elevating from a cigarette. These types of flows have been studied for over 80 years, after Zeldovich's~\cite{zeldovich1937limiting} work on similarity analysis, and Schmidt's~\cite{schmidt1941turbulent}, who obtained analytical expressions for the mean velocity and temperature profile in turbulent plumes. Despite the presence of a  wide variety of experimental~\cite{george1977turbulence,shabbir1994experiments} and theoretical~\cite{morton1959forced} works in this domain, the phenomenology underlying plume physics is still not completely mastered. Numerous works have used CFD to understand this type of flows, based either on Reynolds Averaged Navier-Stokes equations~\cite{van2006application, lou2019numerical} or LES~\cite{zhou2001large}. Even for simplified incompressible and inviscid plumes, the interaction between the gravity source term and the momentum-driven forces results in an interesting and complex behavior. The misalignment of the density and pressure gradients causes the generation of vorticity and the development of instabilities~\cite{turner1979buoyancy}. Thus, even with a simplified setup, a variety of phenomena with different spatial scales can be found. For further information, the reader can refer to the reviews of Turner~\cite{turner1969buoyant} and Hunt et al.~\cite{hunt2011classical}. 

 The plume case has also been widely treated in the computer vision community, due to the difficulty of calculating realistic smokes in real-time, which requires efficient computational solvers, and explains why Machine Learning was first introduced in this context. One can cite the works of Chu et al.~\cite{chu2017data} with a CNN, Kim et al.~\cite{kim2019deep} using GAN with a novel stream function-based loss and Tompson et al.~\cite{Tompson2017} who introduced the idea of a CNN coupled to an incompressible fluid solver.

\begin{figure}[h!]
    \centering
    \includegraphics[width = 0.7 \textwidth]{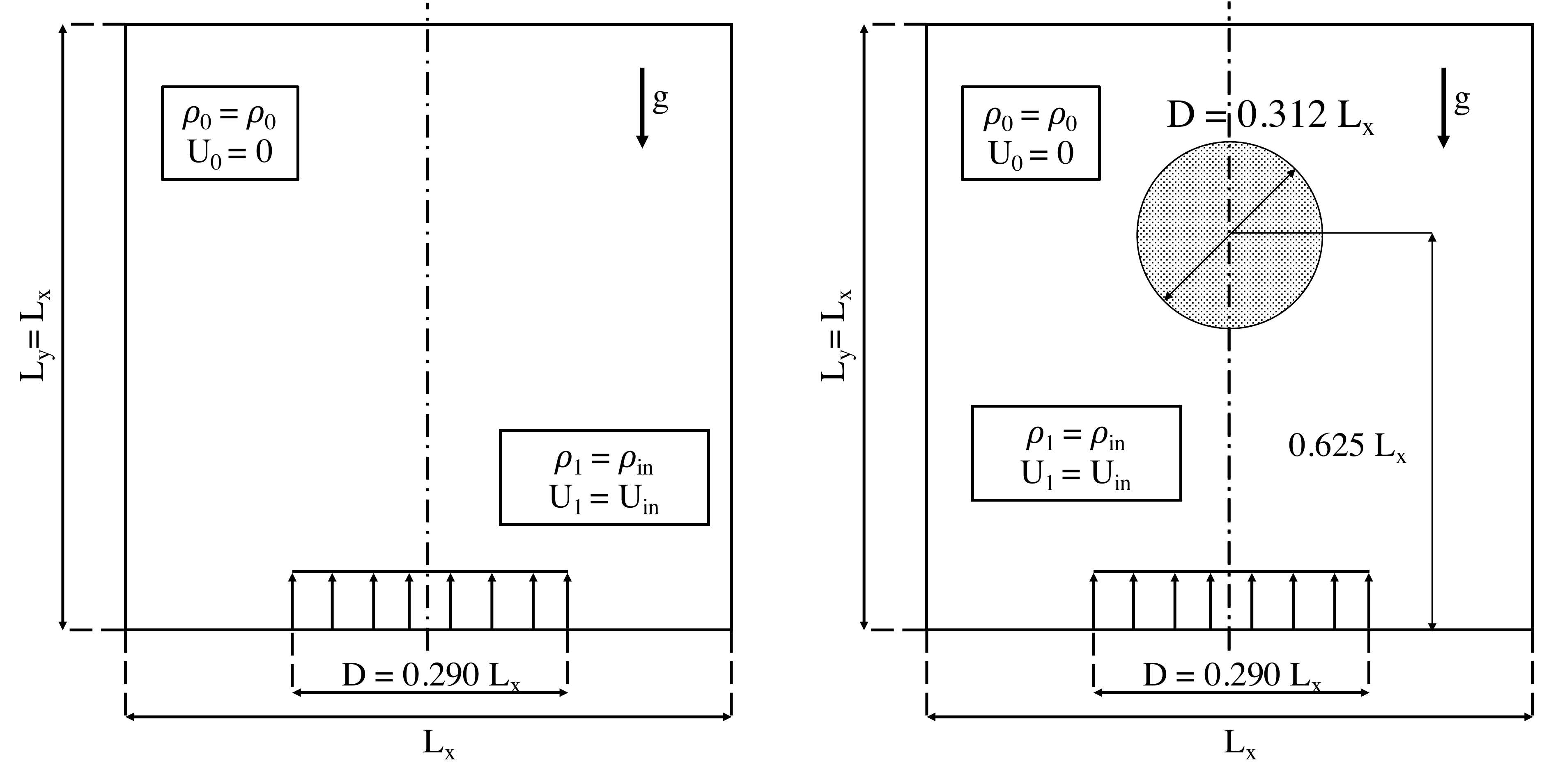}
    \caption{Plume configuration with and without cylinder}
    \label{fig:plume_config}
\end{figure}

The two configurations studied in this work are displayed in Fig.~\ref{fig:plume_config}. The flow is assumed inviscid  (infinite Reynolds number) and adiabatic. Thus, the plume behavior is driven by only one dimensionless number, called the Richardson number (Eq.~\ref{eq:Ri}), which compares the buoyancy and inertia forces:

\begin{equation}
Ri = \frac{ \frac{\Delta \rho}{\rho_{0}} g L}{U^{2}},
\label{eq:Ri}
\end{equation}

\noindent where $\Delta \rho / \rho_{0}$ is the density contrast between the light and heavy fluid, \textit{g} is the gravity, $L$ a characteristic length, and $U$ the velocity of the injected lighter fluid.  High Richardson numbers imply that the flow is piloted by the buoyant forces, whereas low Richardson numbers indicate that the flow is driven by momentum (also known as jets). In both cases, the domain is a square box of length $L_X$, with an inflow centered in the bottom face of the domain, of radius $D_{in} = 0.29L_X$. In the second configuration, a cylinder of diameter $D_c = 0.312L_X$ is added, aligned with the horizontal axis and centered at $0.625L_X$ on the vertical axis (Fig.~\ref{fig:plume_config}). The objective is to evaluate the behavior of the proposed learning-based approach when the flow impinges objects and walls.

\subsection{Numerical setup}
\label{sec:equa}

%{\color{red} si tu fais du Euler, parle simplement du Euler ...  a la limite dit que tu peux sans difficulte faire du NS, mais ensuite met les equations d'Euler sinon on ne comprend pas bien.}

For the CFD study, a fluid solver based on the open-source solver Mantaflow \cite{Thuerey2016} is used to solve the Euler equations. A variable-density incompressible formulation is employed with external forces due to gravity. To model buoyancy-driven flows under the incompressibility assumption,  the Boussinesq approximation~\cite{gray1976validity} is applied, which holds for small density variations compared to the mean density, i.e., $\Delta \rho/ \rho_{0} \ll 1 $.  Note that the continuity equation with the Boussinesq approximation and the incompressibility constraint leads to the advection of the density variation $\Delta \rho$ as a passive scalar, i.e., being constant along a streamline. 

The Euler equations are solved in two steps: (i) the advection step and (ii) the pressure projection step. The computational domain is discretized on a uniform cartesian grid with a finite difference scheme on a Marker-and-Cell (MAC) staggered grid~\cite{harlow1965numerical}. First, the density is convected along streamlines, using an unconditionally stable, second-order accurate Maccormack semi-Lagrangian method~\cite{Selle2008}. Once the density field is advected, the momentum equation is solved through a splitting method, where the velocity field is first advected similarly to the density field, and then modified by the external forces with a forward Euler scheme, resulting in the velocity field $u^{*}$. Since the pressure term in the momentum equation and the mass conservation equations have not yet been used, the velocity field $u^{*}$ does not satisfy the incompressible kinematic constraint $\nabla \cdot u^{*}=0$. To do so, this pressure term is used as a correction term to ensure a divergence-free velocity field, so that the corrected velocity reads:

\begin{equation}
    \textbf{\textit{u}} \simeq \textbf{\textit{u}}^{*} - \Delta t \frac{1}{\rho_{0}} \nabla p
\label{eq:update_Pois}
\end{equation}

\noindent Taking the divergence of both sides of Eq.~(\ref{eq:update_Pois}) and imposing the continuity equation $\nabla \cdot \textbf{\textit{u}} = 0$ yields the well-known Poisson equation:

\begin{equation}
    \frac{ \Delta t}{\rho_{0}}\nabla^{2} p  =  \nabla \cdot \textbf{\textit{u}}^{*}
    \label{eq:Poisson}
\end{equation}

\noindent This equation is solved to obtain the pressure correction, ensuring that the velocity field at the next time-step satisfies both the momentum and the mass conservation equations. This correction is applied using Eq.~(\ref{eq:update_Pois}). 

The Poisson equation is a second-order elliptical partial differential equation, which is well known to be computationally expensive, taking up to $80\%$ of the computational time of the incompressible solver~\cite{ajuria2020towards}. For this reason, this work  only focuses on the resolution of this particular equation. This equation may be solved using direct methods~\cite{buzbee1970direct,quarteroni2010}, such as the LU factorization, which can be interpreted as a Gaussian Elimination Method (GEM). The idea behind these methods is to recast the A matrix of the linear system $Ax =b$  into an upper triangular and a lower triangular matrix $L U x =b$, which enables to decompose the problem into two successive GEMs, $Ly =b$ and $Ux =y$ respectively. These methods can provide solutions at the machine-precision, however, they tend to be computationally expensive, especially for large computational domains since the cost of solving a single GEM is proportional to $2n^3/3$ operations, where $n$ is the number of points in the mesh. Iterative methods have been proposed to avoid the direct solution of the system. For such methods, a trade-off has to be chosen between accuracy and computational cost. Despite the need of multiple iterations to achieve convergence, those methods are usually more efficient than direct approaches since their computational cost is proportional to $n^2$ per iteration. Among the iterative methods, the Jacobi method, which was first developed in 1845~\cite{jacobi1845}, by the German mathematician Carl Gustav Jacobi, decomposes the matrix A into a diagonal matrix D and the upper and lower diagonal matrices L and U. The linear system can be expressed as $D X = b - (L+U)X$, which is equivalent to  $X = D^{-1} (b - (L+U)X)$. An iterative process is obtained by solving the previous equation in the form $X_{t+1} = D^{-1} (b - (L+U)X_t$. As any iterative method, the number of iterations required towards convergence strongly depends on the initial guess $X_0$: an additional result of the present work compared with previous studies is to use the neural network prediction as initial guess to speed-up the Jacobi method. More advanced methods~\cite{saad2003,hosseinverdi2018efficient} exist for the Poisson equation, yet this work will focus only on the Jacobi method for the sake of stability, its easy implementation on GPUs, as well as its straightforward hybridisation with neural networks.

\section{A deep-learning approach to solve the Poisson equation}
\label{sec:DL_poisson}

To overcome the difficulties of iterative methods, the present study focuses on solving the Poisson equation with a Convolutional Neural Network (CNN). Particular attention is put on its ability to tackle new cases, not seen during the training phase. To further ensure both a fast and reliable prediction of the corrected velocity field, this CNN method will be hybridized with the Jacobi method, using the CNN prediction as initial guess to significantly speed-up the Jacobi approach.

Convolutional Neural Networks perform multiple linear transformations, typically $y = w_{i} x + b_{i}$, followed by non-linear activation functions to output the desired field. The \textit{training} of the neural network consists in the optimization of the weights $w_{i}$ and biases $b_{i}$ to minimize a user-defined objective function $\mathcal{L}$. CNNs differ from traditional MLP as they do not apply the weights and biases to the entire input fields. Indeed, CNNs apply the transformation locally through a kernel of limited size, typically $3 \times 3$, which is then `slid' through the entire input field. This type of network is widely used for image recognition tasks, as the kernels can be interpreted as filters that extract patterns and structures from the input images. CNNs are therefore particularly adapted to fluid mechanics, since the underlying physics are mostly driven by flow structures (jets, vortices, stagnation points etc.). Additionally, compared with MLP, CNNs usually require a smaller number of tunable parameters ($w_i$ and $b_i$), and are theoretically independent of the size of the input field: the same trained network can be reused on cases with different resolutions or meshes. However, a current limitation of CNNs is the need of perfectly uniform cartesian grids, whereas most of today CFD simulations are performed on non-uniform, possibly unstructured, meshes. While interpolation from non-uniform meshes onto a uniform cartesian grid is possible, recent works have efficiently adapted convolution operations to unstructured meshes~\cite{baque2018geodesic}, making CNNs even more suitable for CFD applications. However, these methods are still complex and not fully mature, so CNNs are still widely used on CFD. For the sake of simplicity, this study considers a perfectly uniform cartesian grid. The spatial resolution will however be varied to validate the convergence properties of the method, as well as to evaluate its computational performance on different grid sizes.

\subsection{Network architectures}
\label{sec:network_architecture}

The choice of the neural network architecture is usually driven by the user's experience, and further improved by a trial and error approach. Since the network architecture is critical, tools for automatic architecture search have been also developed~\cite{elsken2018neural}. While discovering efficient network architectures, these automated data-driven search strategies require an extremely high computational time since multiple trainings have to be performed, as shown by Zoph et al.~\cite{zoph2018learning} who used in parallel 500 Nvidia p100 GPUs during 4 days to find an optimal network for scalable image classification. Architectures are even more critical in physics, and especially fluid mechanics, since the physics usually involve a large range of different scales which need to be captured accurately by the network. Towards this objective, Geng et al.~\cite{geng2020automated} have employed such method to discover optimal neural networks dedicated to multi-scale analysis of seismic waves. However, despite those automatic strategies, the choice of the network architecture and its effect on accuracy and inference time is often ignored. In particular, guidelines still need to be established for physics-related tasks. Here, different types of network architectures are investigated, in order to analyze their effect on the resolution of the Poisson equation. Note that the objective of this work is not to find the optimum architecture for the resolution of the Poisson equation, but rather to provide general trends and understanding of the architecture effect on the network accuracy and performance. For all architectures tested here, the pressure field is computed using a CNN for which the inputs are the uncorrected velocity divergence $\nabla \cdot u^*$ and a boolean field describing the object geometry. 

\begin{figure}[h!]
    \centering
    \includegraphics[width = 0.7 \textwidth]{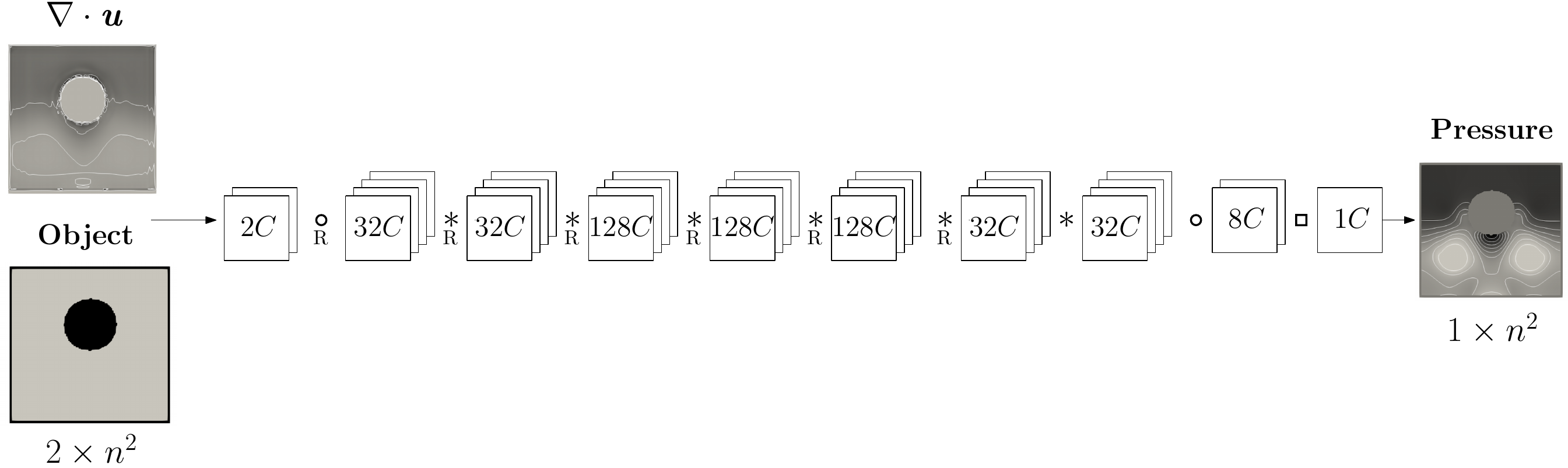}
    \caption{MonoScale network containing  395601 parameters, where $\circ$ corresponds to convolution with kernels of size 5$\times$5, $\ast$ to kernels of size 3$\times$3 and $\square$ to kernels of size 1$\times$1. R corresponds to the ReLU activation function. Each box indicates the number of feature maps, or channels ($C$) present at each layer.}
    \label{fig:Mono}
\end{figure}

First, a simple "feed-forward" convolutional network is considered, denoted hereafter MonoScale. This network (Fig.~\ref{fig:Mono}) contains $395601$ tunable parameters distributed in 9 layers, following a straightforward architecture. For convolution operations, a replication padding is used to ensure that all the feature maps keep the size of the original fields. This type of padding creates \textit{ghost cells} on the image boundaries with a value copied from the closest image pixel. This network is the most simple strategy to tackle the Poisson problem, as the input is just passed through a series of convolutional layers, without any further modification or post-processing.

\begin{figure}[h!]
    \centering
    \includegraphics[width = 0.7 \textwidth]{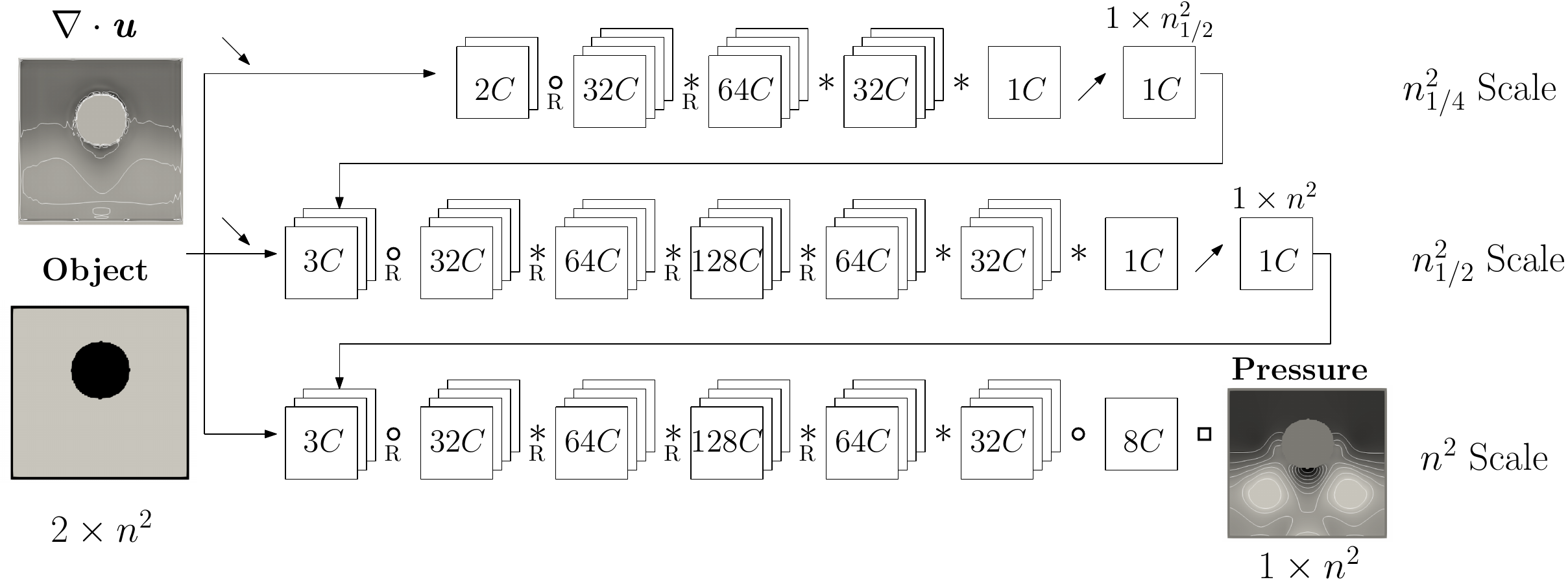}
    \caption{MultiScale network, with  418640 parameters, where $\circ$ corresponds to convolution with kernels of size 5$\times$5, $\ast$ to kernels of size 3$\times$3 and $\square$ to kernels of size 1$\times$1. R corresponds to the ReLU activation function, $\searrow$ indicates a bilinear downsampling operation, whereas $\nearrow$ corresponds to the bilinear interpolation. Each box indicates the number of feature maps, or channels ($C$) present at each layer.}
    \label{fig:MS_Baseline}
\end{figure}

The second architecture is the MultiScale (Fig.~\ref{fig:MS_Baseline}), introduced by~Mathieu et al.~\cite{Mathieu2015} for video image prediction, and introduced to solve the Poisson equation by Ajuria et al.~\cite{ajuria2020towards}. It has also been employed in other studies on fluid mechanics, such as in Fukami et al.~\cite{fukami2018super, fukami2020assessment} focusing on the super-resolution of turbulent flows, or applied to the propagation of acoustic waves by Alguacil et al.~\cite{alguacil2020predicting}. The idea behind this architecture is to feed the neural network with several representations of the same inputs, focusing on different scales. To do so, the original input is interpolated on coarser meshes: while the largest resolution retains all scales, smaller resolutions focus only on the large scales. The network then encodes information linked to the different spatial scales at each level, which helps the network on the generalization task, and avoids spurious high-frequency oscillations on the large features of the image~\cite{Mathieu2015}. Here the MultiScale architecture (Fig.~\ref{fig:MS_Baseline}) has three scales, of sizes $n_{1/4}^2$,  $n_{1/2}^2$ and  $n^2$ respectively. The first scale interpolates the original images on a quarter-cell size mesh. The resulting image is then interpolated on a half-cell size mesh to twice its size ($n_{1/2}^2$). The middle scale takes as an input both the interpolation of the original entries on a half-cell size mesh, and the output of the previous scale. Finally, the resulting image is interpolated to the original size $n^2$, and is concatenated to the input fields. It is then fed to the final scale branch. The first scale contains 5 layers, while both the intermediate and final scales contain 7 layers. Non-linear ReLU activation functions are placed after each convolution layer, except on the last two ones, to enable the network to compute positive and negative outputs. 

\begin{figure}[h!]
    \centering
    \includegraphics[width = 0.9 \textwidth]{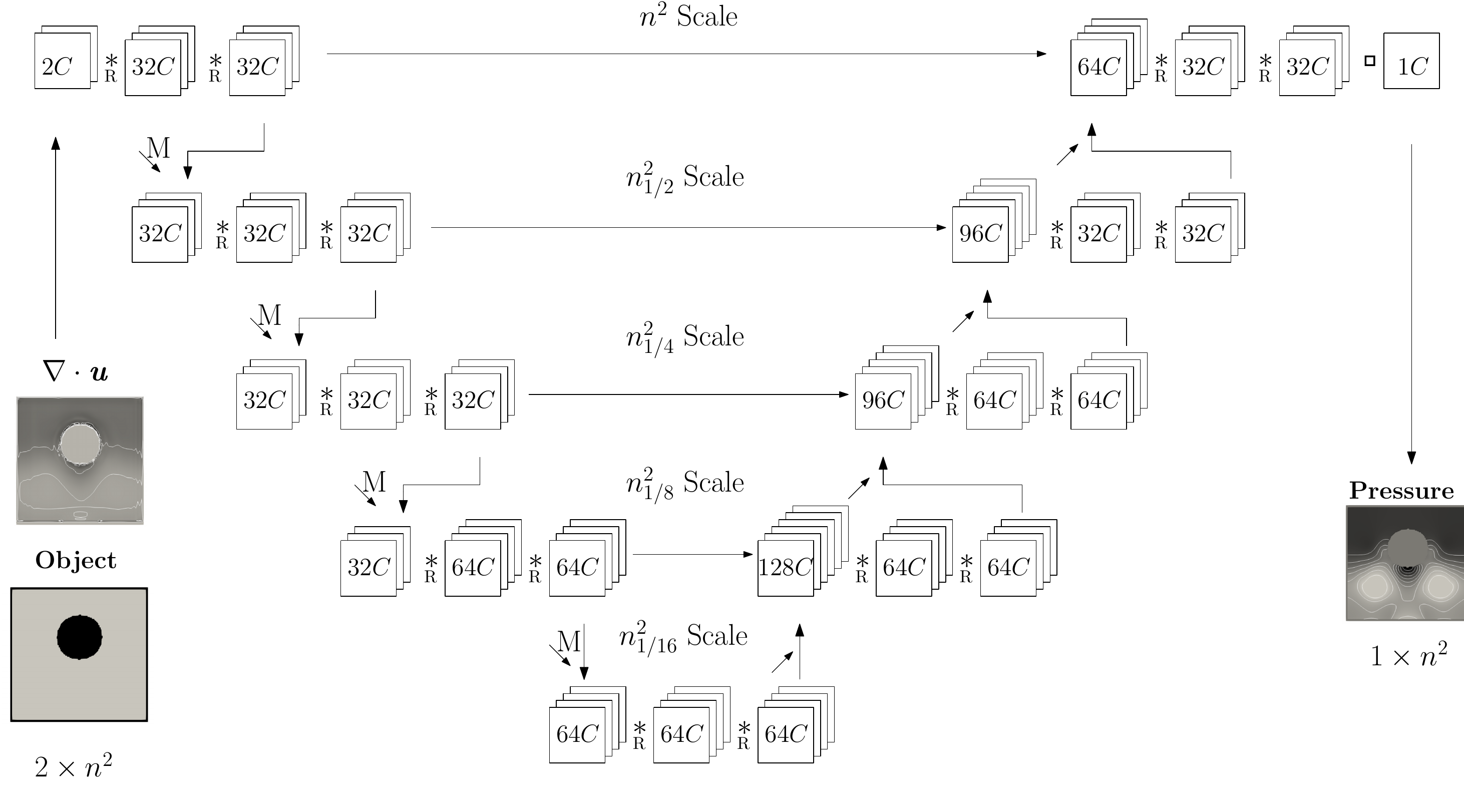}
    \caption{Unet with 443521 parameters, where $\ast$ corresponds to kernels of size 3$\times$3 and $\square$ to kernels of size 1$\times$1. $R$ corresponds to the ReLU activation function, $\searrow$ M indicates a MaxP Pooling operation, whereas $\nearrow$ corresponds to the bilinear interpolation. At each scale, the MaxPooling step reduces the images to 1/2 of its original size, whereas the interpolation upsamples the image to the double of its size. Each box indicates the number of feature maps, or channels ($C$) present at each layer.}
    \label{fig:Unet}
\end{figure}

The MonoScale and MultiScale networks are then compared with the well-known Unet architecture (Fig.~\ref{fig:Unet}), which was first developed by Ronneberger et al.~\cite{ronneberger2015u} for the segmentation of biomedical images. Apart from the initial biology-related applications, this network has also been used in many other fields, including for regression tasks in fluid mechanics. For instance, Lapeyre et al.~\cite{lapeyre2019training} employed this architecture on turbulent sub-grid scale modeling for reacting flows. The structure of the Unet is similar to the MultiScale, as it also combines information extracted at different spatial scales. However, the Unet differs from MultiScale on the scale treatment: while several scales are managed in parallel in the MultiScale network and the input image is fed directly to each branch, the Unet acts as a simple feed-forward network with a decreasing-increasing resolution to encode information in a low-dimensional manifold (at the smaller scale). The image input is fed to the network only at the initial resolution. Additionally, skip connections are imposed at each scale, allowing information to by-pass the lower-resolution treatments and to avoid the vanishing gradient issue inherent to deep network architectures. 

%{\color{red} Il faut que tu sois sur de tout ce que tu dis dans le paragraphe qui suit ! et il faut plus de citations !}
Note that those architectures are similar to multigrid and multi-fidelity approaches developed in CFD. Indeed, the MultiScale network corresponds to multi-fidelity approaches~\cite{mccormick1989multilevel}, where the original problem is solved on a coarse grid, the solution is then used as an initial guess to a more refined grid~\cite{southwell1956relaxation}. The Unet on the other hand is similar to multigrid solvers~\cite{wesseling1995introduction}, where the initial problem is solved with a computationally non-expensive solver, and the result is corrected with approximations computed on coarser grids.

In comparison with classical deep neural networks employed for classification tasks which may contain $10^6$ (e.g. ResNets~\cite{he2016deep}) to $10^8$ (e.g. VGG-16~\cite{simonyan2014very}) tunable weights, the proposed architectures are relatively small, containing all about $4~10^5$ parameters. Since the main goal of the present approach is to accelerate the classical iterative Poisson solvers, a limited number of parameters have been chosen. Note however that this cannot constitute a guideline since the performance of the networks will be evaluated at iso-level of accuracy: the number of parameters is, therefore, a non-trivial trade-off between accuracy and inference time, for which no rule exists in the literature. To investigate this accuracy-performance link, a smaller MultiScale containing $1.3~10^5$ parameters is introduced in complement to the three previous networks. The structure is exactly the same as the MultiScale, but only the number of filters per layer is changed, resulting in a network with around 3 times fewer parameters. This network will be denoted \textit{SmallScale} in the following.

%{\color{red} Rajoute-on le MultiScale a 5 niveaux ou le Unet à 3 niveaux avec 400k parametres pour donner une indication de l'effet de ces echelles a iso-nombre de param?}

Consequently, four deep neural networks will be analyzed on the two test cases introduced in Section~\ref{sec:plume_cases}. For each architecture, both their error levels and distributions, as well as their performances on inference time, will be assessed. 

\subsection{Training and Loss function}

The networks introduced in Section~\ref{sec:network_architecture} are trained using a procedure similar to Tompson et al.~\cite{Tompson2017} and Ajuria et al.~\cite{ajuria2020towards}. Compared with most studies on Machine Learning using a known output as target in the loss function $\mathcal{L}$ for a supervised training~\cite{lecun1998gradient}, here a semi-supervised learning strategy is used, where no `ground truth' field is needed. To do so, the residual of the continuity equation is used as loss function, where the divergence of the velocity field is computed using a first-order forward finite difference scheme. Thus, training the network is equivalent to minimizing this residual, i.e., to enforce the mass conservation equation, which is the goal of solving the Poisson equation in an incompressible solver. Note that this approach can be considered as a Physics-Driven Neural Network (PDNN), demonstrating how a target physical equation can be introduced into the learning strategy. It differs from Physics-Informed Neural Networks (PINN,~\cite{Raissi2017}), which combines a semi-supervised physics-driven approach with a standard supervised learning method in which a 'ground truth' target is required. The main benefit of PDNN over PINN, is the possibility of adding long-term loss effects during the training by constraining the network to produce consistent predictions in time.

\begin{figure}[h!]
    \centering
    \includegraphics[width=0.8\textwidth]{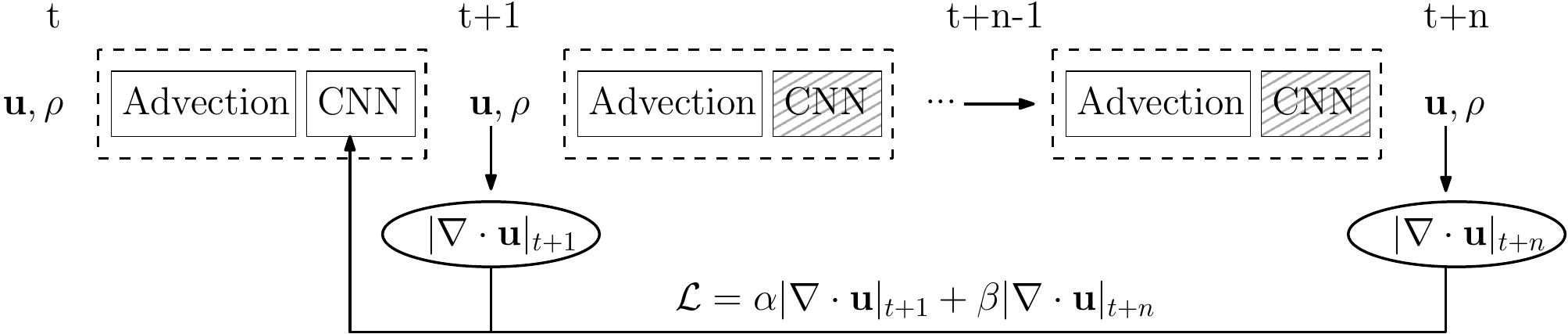}
    \caption{PDNN learning strategy by combining a \textit{short-term} and a \textit{long-term} loss. The tiled box for the CNN indicates that the network parameters are frozen (i.e. they are the same as the one used in the network at time $t$).}
    \label{fig:loss_lt}
\end{figure}

To train the network, the widely-used Mean Squared Error (MSE) metric~\cite{hunt1973application} is introduced, which computes the L2 distance between two fields. As the loss function of the PDNN is the residual of the continuity equation, the MSE metric is applied directly to the divergence of the corrected velocity, which can be interpreted as the L2 distance to zero. Since the solution of the Poisson equation is the pressure correction term, the network output has to be coupled with the fluid solver to produce the desired corrected velocity and compute its divergence and L2 norm. Note that the gradient of this correction (Eq.~\ref{eq:update_Pois}) is known, so that a classical backpropagation can still be employed. However, if the network is only trained by evaluating the divergence level after a single timestep, no information is given to the network on how its error can be amplified by the non-linear advection at the next timestep. To circumvent this issue, a new term is introduced into the PDNN loss. This new term, known as long-term loss, corresponds to the divergence field of the uncorrected velocity field obtained several time steps after the initial prediction (Fig.~\ref{fig:loss_lt}). To do so, the network is trained inside the CFD solver: the prediction of the CNN at time t is fed through the advection and the CNN-based Poisson solver to compute the next n timesteps. The resulting divergence is computed (\textit{long-term} loss in Fig.~\ref{fig:loss_lt}), and added to the initial divergence (\textit{short-term} loss). The network weights are then updated using a gradient descent algorithm. Note that in the backpropagation algorithm, the chain rule~\cite{hecht1992theory} is used, yet here the advection part is not differentiated, and therefore does not appear explicitly in the gradients. The number $n$ of time steps to compute the long-term divergence term is either $4$ ($90\%$ of the times) or $16$ ($10\%$ of the times). As a summary, the total loss function is:

\begin{equation}
\mathcal{L} = \frac{\alpha}{N} \sum^N \left|\nabla \cdot {\bf u_{t+1}} \right| + \frac{\beta}{N} \sum^N \left|\nabla \cdot {\bf u_{t+n}} \right|
\end{equation}
where $\alpha$ and $\beta$ are two hyperparameters controlling the relative importance of the short-term and long-term losses.

As a semi-supervised training procedure is considered, a complete training dataset is not necessary to train the network since no 'ground truth' target is needed. However, physical initial conditions are still possible, in particular to enforce the spatial coherence and structures relevant from a physical point of view, and to avoid overfitting. Thus, even for this PDNN, a training dataset was computed with the open-source code Mantaflow~\cite{Thuerey2016}. This dataset consists of closed boundary domains, with randomly initialized velocity fields and no density variations (i.e. the Richardson number is $R_i = 0$ for all training examples). Random geometries are placed in arbitrary locations, and velocity divergence sources are introduced in the domain to initialize the flow motion. As a result, the dataset consists of $320$ simulations on a $128 \times 128$ uniform cartesian grid. Each simulation is run for 64-time steps. 

\subsection{Hybrid Methodology}
\label{sec:hybrid_methodology}

\begin{figure}[h!]
    \centering
    \includegraphics[width=0.7\textwidth]{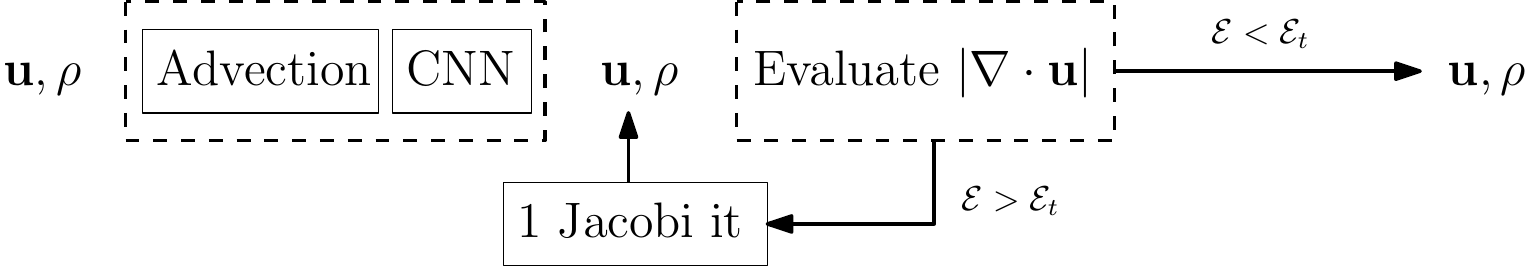}
    \caption{Sketch of the hybrid method, which is activated depending on the error $\mathcal{E}$ compared with the threshold value $\mathcal{E}_t$}. %{\color{red} Remplace tol par $\mathcal{E}_t$}.}
    \label{fig:hyb_scheme}
\end{figure}

The main difficulty when building a Machine Learning-based CFD solver is to guarantee its precision and robustness, in particular for cases far from the learning database. To do so, Ajuria et al.~\cite{ajuria2020towards} introduced a hybrid strategy, where the network error is tracked over time. Depending on this error level, the network prediction is used as an initial guess for the Jacobi method to further improve its quality. Note that a similar procedure was also proposed by~Hsieh et al.~\cite{hsieh2019learning}, where an iterative solver is coupled to a deep linear network to guarantee the accuracy level of the network prediction. An overview of the hybrid method is shown in Fig.~\ref{fig:hyb_scheme}: after the advection step, the pressure correction term is predicted by the deep neural network architectures proposed in Section~\ref{sec:network_architecture}. At this point, no guarantee is given on the precision of the method, possibly leading to large errors or even numerical instabilities~\cite{ajuria2020towards}. The hybrid strategy is then applied, based on an error associated with the divergence velocity field, denoted $\mathcal{E}_{\infty}$. Ajuria et al.~\cite{ajuria2020towards} proposed to use the maximum of the divergence velocity field, i.e. $\mathcal{E}_{\infty} = max |\nabla \cdot {\bf u}|$, showing a significant improvement of the solver accuracy and robustness. Note however that no physical evidence of this choice is given. In particular, there is no formal proof that reducing the maximum of the velocity divergence actually leads to a more physical behavior of the simulation. One objective of this paper is to propose a choice of $\mathcal{E}$ that guarantees a physical behavior of the simulation, towards a fast, reliable and robust CNN-based incompressible solver.

\section{Error Quantification}
\label{sec:errorquant}

In order to compare the performance of the several architectures defined in Section~\ref{sec:network_architecture}, a detailed analysis of the network errors is required. Indeed, evaluating and comparing the time of inference of a method, in particular with neural networks, is relevant only if performed at a fixed error level. Thus, this section intends to characterize the accuracy of each architecture, so that a fair comparison of performances can be achieved in Section~\ref{sec:performance}.

\subsection{Preliminary results without hybrid approach}
\label{sec:no_hybrid}

First, the four networks described in Section~\ref{sec:network_architecture} are tested without the hybrid approach on a plume test case with a Richardson number $R_i = 14.8$, in order to evaluate the network accuracy and its generalization capabilities (since the training dataset was obtained for $R_i = 0$). As the Boussinesq approximation is used, the density variation should remain small and is set here to $\Delta \rho / \rho_{0} = 0.01$. The lighter fluid is injected at the inlet boundary with the velocity $U_{in} = 0.01 m s^{-1}$. The CFD domain size is $L_X = L = 512m$, discretized on a uniform cartesian grid of $512 \times 512$ cells. The gravity is set to $g = 0.002 m s^{-2}$, and the inlet radius is $R = 74,25m$, taken as the characteristic length to define the Richardson number. The divergence is normalized with the inlet velocity and the characteristic length, and time is normalized as well by the characteristic time needed for the plume to reach the top boundary at $R_i = \infty$, i.e., in the pure jet configuration with no buoyancy. These normalized quantities are therefore defined as

\begin{equation}
\widetilde{u} =  R~u / U_{in} ~\mbox{ and }~ \tilde{t} = t~U_{in} / L
\end{equation}

\begin{figure}[h!]
    \centering
    \includegraphics[width=0.5\textwidth]{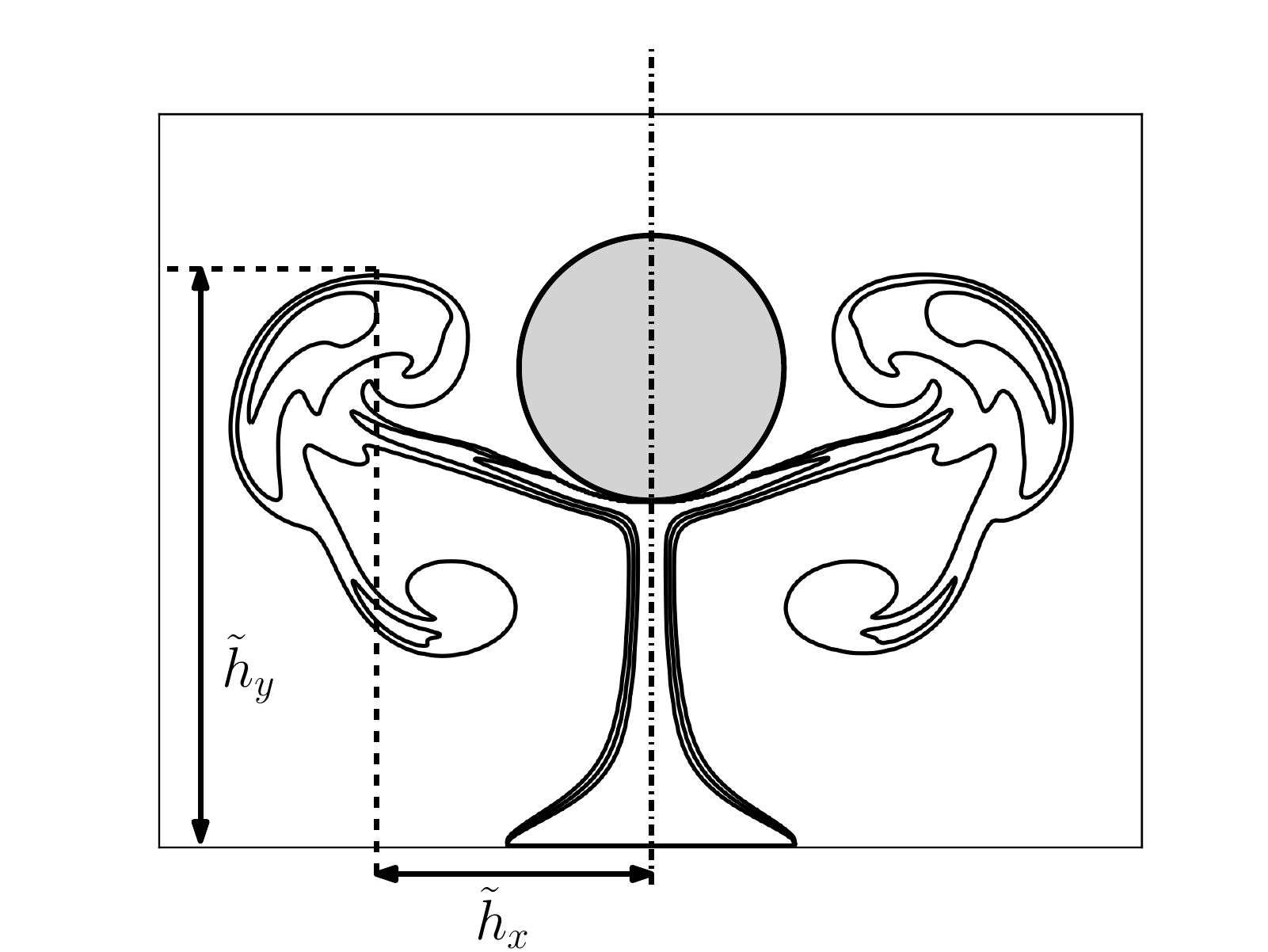}
    \caption{Sketch of the plume-cylinder configuration. $\tilde{h}_x$ and $\tilde{h}_y$ are the coordinates of the plume head location.}
    \label{fig:cyl_scheme}
\end{figure}

The plume head position is defined as the highest point of the plume at a time $t$ (Fig.~\ref{fig:cyl_scheme}). On the vertical axis, this position is noted $\tilde{h}_y = h_y / L $. In the case of the plume impinging the cylindrical obstacle, the plume head position along the horizontal axis is also measured, denoted by $\tilde{h}_x = h_x / L $ (Fig.~\ref{fig:cyl_scheme}). The vertical position $\tilde{h}_y$ allows the identification and comparison of the plume velocity when rising due to both the advection by the jet and the buoyancy forces. The horizontal position $\tilde{h}_x$ measures the flow deviation induced by the cylinder. This measurement is performed in both the left and right sides of the CFD domain in order to quantify the asymmetry of the flow field. 

\begin{figure}[h!]
    \centering
    \includegraphics[width=0.95\textwidth]{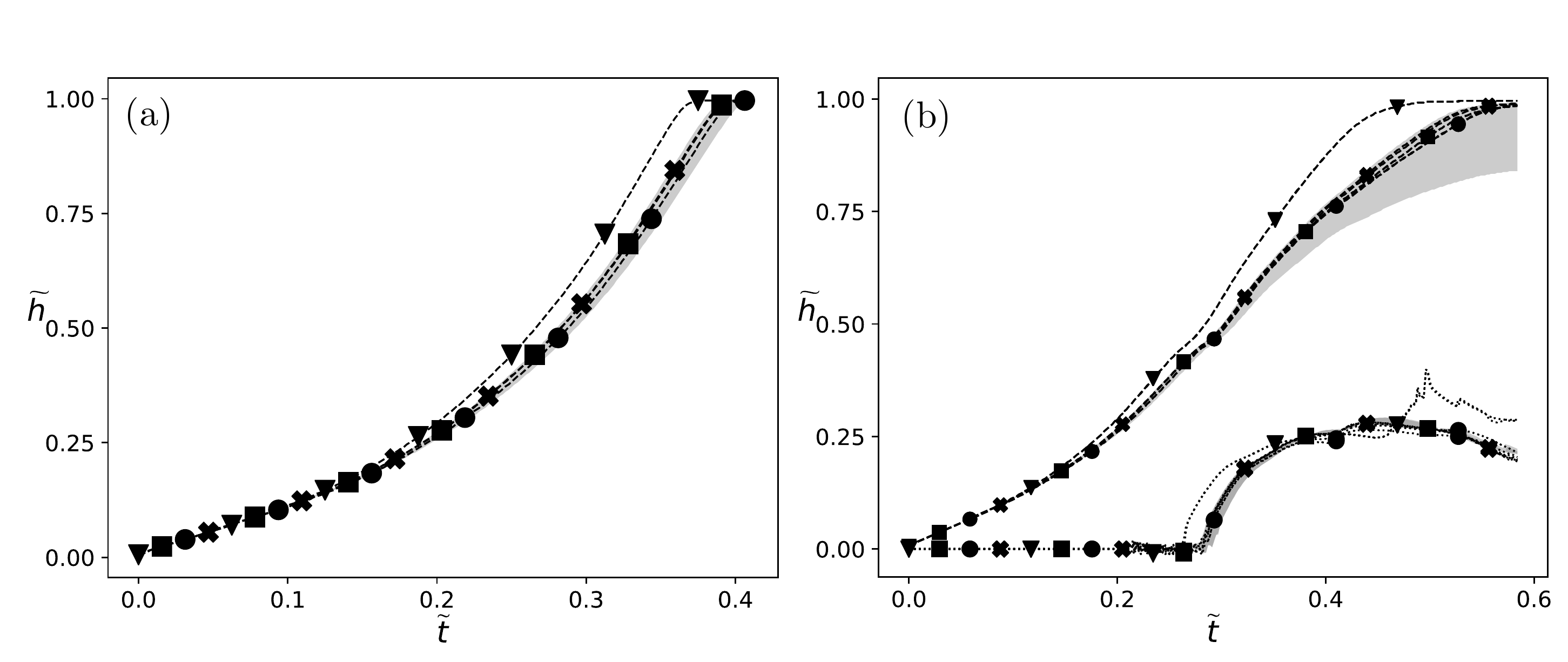}
    \caption{Plume head coordinates $\tilde{h}_x$ (\dotted) and $\tilde{h}_y$ (\dashed) for the case without (a) and with (b) obstacle at $R_i=14.8$ obtained by several networks: $\blacktriangledown$ MonoScale, $\blacksquare$ MultiScale, $\bullet$ Unet, and $\times$ SmallScale. The grey zone shows the range of the plume head position obtained with the Jacobi solver, using between 200 and 10000 iterations.}
    \label{fig:Basic_head}
\end{figure}

Analyzing the plume head's position for the several networks on the two test cases presented in Fig.~\ref{fig:plume_config} reveals that deep neural networks are able to solve the Poisson equation in an Euler incompressible solver with good precision, even in cases not seen during training (Fig.~\ref{fig:Basic_head}, tested at $R_i = 14.8$). However, slightly different plume developments are obtained depending on the network architecture. As introduced in previous work~\cite{ajuria2020towards}, the Jacobi solver is taken as the reference for the two studied test cases. The accuracy of the Jacobi solver solution increases with the number of iterations, but as the convergence rate does not follow a linear behavior, a trade-off between the number of iterations and the desired accuracy is usually necessary. The grey zone in Fig.~\ref{fig:Basic_head} represents the solution range obtained with between 200 and 10000 iterations. Note that the plume development is slower with higher number of iterations. While the solution deviation of the test case without obstacle remains narrow, the cylinder test case shows a larger dispersion. Due to computational cost, it was however not possible to perform more than 10000 iterations of the Jacobi solver.

%{\color{red} Il manque peut etre un Jacobi faible et eleve pour comparaison e tservir de reference}. 

\begin{figure}[h!]
    \centering
    \includegraphics[width=0.95\textwidth]{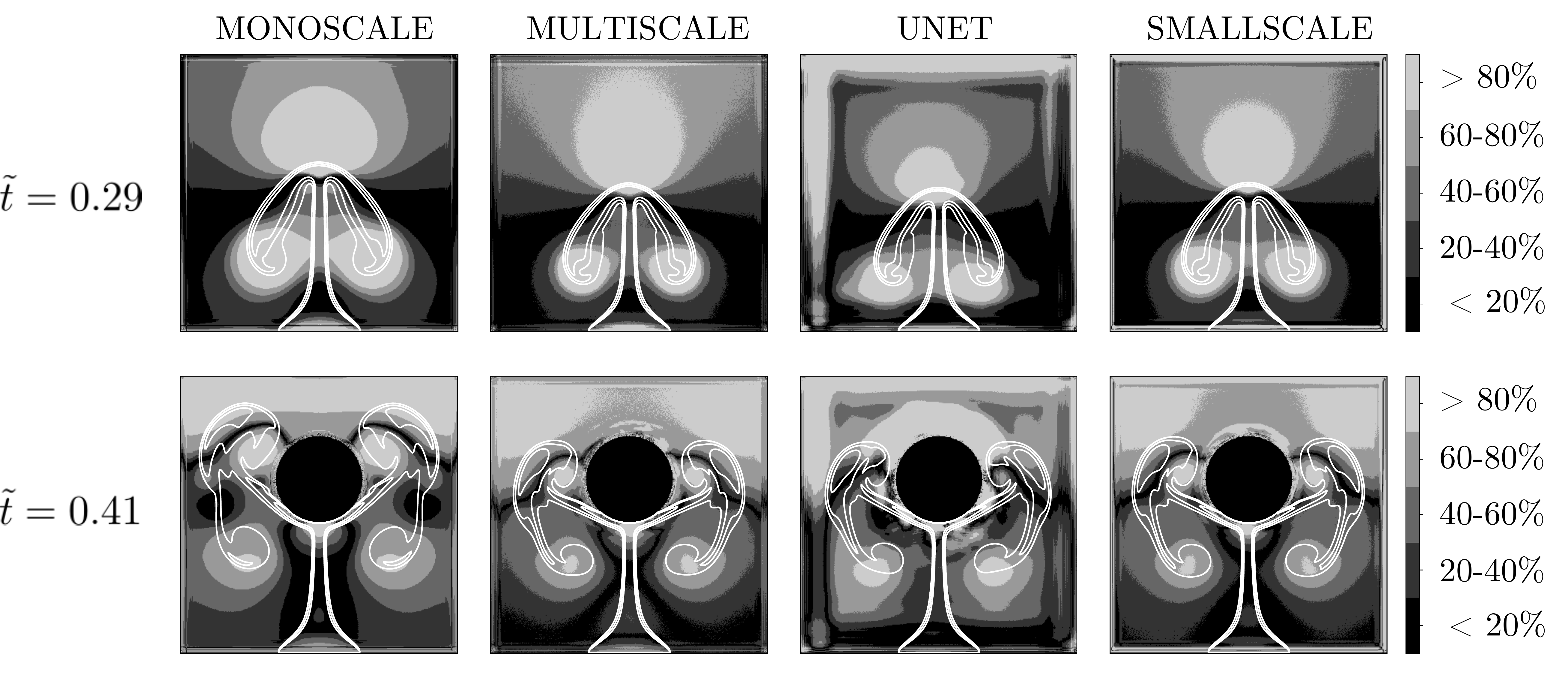}
    \caption{Divergence error percentiles and density iso-contours (in white) of the four studied networks, at time $\widetilde{t} = 0.29$ for the case with no cylinder (top row) and time $\widetilde{t} = 0.41$ for the cylinder case (bottom row).}
    \label{fig:heat_map_snapshots}
\end{figure}

For the no-cylinder case, the MonoScale network results in a faster head propagation, which falls outside the confidence interval of the Jacobi solvers, whereas all other network architectures produce an overall good physical behavior of the plume. The same trend is observed in the case with the obstacle, with a faster propagation and an early flow deviation for the MonoScale. Note that the symmetry is preserved by all networks since no difference is observed between left and right horizontal positions $\tilde{h}_x$. The final horizontal length of MonoScale does not match the ones of the other networks. This is due to a wrong prediction of the flow structure and roll-up of the plume by the MonoScale network. This is highlighted in Fig.~\ref{fig:heat_map_snapshots}, showing the density iso-contours (white lines) and the percentiles of the divergence error  (gray background field) at $\tilde{t} = 0.29$ (no-cylinder case) and $\tilde{t} = 0.41$ (case with obstacle). A significant difference of the plume shape is observed between the MonoScale and the other networks. The MonoScale network predicts larger vortices, probably due to an inaccurate pressure field prediction which negatively affects the baroclinic torque. However, if the divergence error is analyzed, the MonoScale follows the most regular distribution, whereas the rest of the networks seem to struggle in the surroundings of the cylinder. Moreover, the Unet network shows a quite irregular divergence distribution, with problems at boundaries. The complete evolution of the plume for both cases are provided in~\ref{appendix:heatmaps} (Figs.~\ref{fig:initial_field_nocyl} and~\ref{fig:initial_field_cyl}).

In order to provide robust and reliable CNN-based incompressible solvers,~Ajuria et al.~\cite{ajuria2020towards} proposed to track the network error in time, and introduced a hybrid strategy to improve the network predictions if this error level becomes unacceptable. Previously, the head location has been employed for measuring the error, yet it is case-dependent and requires a reference simulation, which prevents this type of error to be used in new unknown simulations. The percentiles of the divergence error displayed in Fig.~\ref{fig:heat_map_snapshots} highlight the locations in the flow field where the neural networks have difficulties, but it cannot be used as an absolute error level as well. Thus, the objective is to define an absolute unsupervised error $\mathcal{E}$, which can be evaluated for any plume case without reference, and is consistent with the physical-based error given by the head position. Here, two measures are considered: (i) the maximum of the divergence velocity field $\mathcal{E}_{\infty}= max(|\nabla \cdot {\bf u}|)$, and (ii) the mean of this divergence field, $\mathcal{E}_{1} = mean(|\nabla \cdot {\bf u}|)$. This type of error control is typically used in iterative algorithms, such as the Jacobi method or the Conjugate Gradient method~\cite{hestenes1952methods}. This assumes a uniform error distribution, which implies that the maximum or mean error value is representative of the overall flow behavior. However, this hypothesis does not necessarily hold for neural networks, since their error distribution is usually non-uniform: neural networks may lead to a very high localized error, near a boundary condition for example. 

\begin{figure}[h!]
    \centering
    \includegraphics[width=0.99\textwidth]{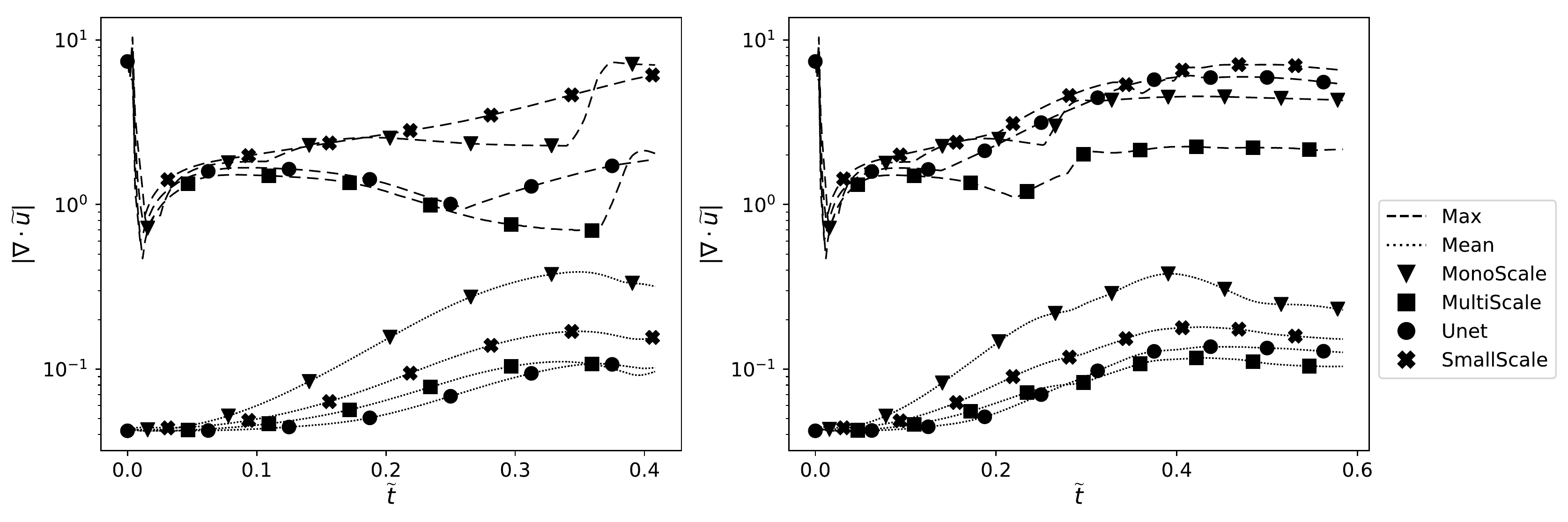}
    \caption{$\mathcal{E}_{\infty}$ (\dashed) and $\mathcal{E}_{1}$ (\dotted) for the case without (a) and with (b) cylinder obtained by several architectures: $\blacktriangledown$ MonoScale, $\blacksquare$ MultiScale, $\bullet$ Unet, and $\times$ SmallScale.}
    \label{fig:Basic_div}
\end{figure}

Figure~\ref{fig:Basic_div} shows these two errors $\mathcal{E}_{1}$ and $\mathcal{E}_{\infty}$ for both test cases without (a) and with (b) obstacle, for the various network architectures. Interestingly, the `best' network is different for the two types of error. With the error based on $\mathcal{E}_{\infty}$, the MultiScale network outperforms the other architectures, in particular for the case with cylinder, whereas the SmallScale leads to the highest error. Regarding $\mathcal{E}_{1}$, the Unet gives a low error in both cases, similar to the MultiScale network. However, based on this error, this is the MonoScale which is the worst network in both situations, with the highest mean divergence. These test cases reveal how critical is the choice of the absolute error $\mathcal{E}$ when comparing and then choosing network architectures.

%The error propagation analysis is a longstanding field, where many works have studied the theoretical propagation of the mathematical error~\cite{pang1997error}, or the impact on the mathematical error on flow topologies, such as on the 3D Taylor Green Vortex~\cite{brachet1983small}. For neural networks, recent works started to further analyze the correlation and behavior of the feature maps inside the neural network, such as developing a Singular Vector Canonical Correlation Analysis on the feature maps of a neural network solving the Poisson equation~\cite{magill2018neural}, or analyzing the feature map content on a 3D Von Karman Vortex shedding~\cite{lee2019mechanisms}. However, some further work is still needed, as the majority of the effort concerning the neural network interpretation focuses on image classification tasks~\cite{montavon2018methods, lundberg2017unified}, analyzing the impact of different feature maps on the final output class. However, on regression tasks where the output field is a continuous field, tracking the origin of different flow structures still needs further development.

%Thus, this work tries to understand which metric would be the most beneficial for the hybrid strategy, in order to ensure a homogeneous flow behavior, particularly on the plume test case with and without a cylinder on the fluid domain. Particularly, a threshold for the mean and for the maximum divergence level will be analyzed.

\subsection{Controlling the error level using the hybrid approach}
\label{sec:control_error}

%The previous section highlighted how critical is the choice of the error definition to compare network architecture. Thus, since the several networks have different accuracy, it is not possible to perform a fair comparison of performance: accelerating a CFD solver is only relevant at a fixed error-level. To circumvent this issue, the hybrid approach introduced in Section~\ref{sec:hybrid_methodology} will be used to control in time the error of each network. Yet, two main problems remain:
%\begin{itemize}
%    \item the two definitions (mean and maximum error level) yield different conclusions on the networks accuracy. 
%    \item up to now, no link exists between these divergence errors and the physical targets $\tilde{h}_x$ and $\tilde{h}_y$.
%\end{itemize}

Before assessing the performance of each neural network in Section~\ref{sec:performance}, it is essential to ensure that the hybrid approach is able to control the physical targets $\tilde{h}_x$ and $\tilde{h}_y$. To do so, this Section intends to determine if $\mathcal{E}_{1}$ or $\mathcal{E}_{\infty}$ is able to control the physical behavior of the simulation. In other words, a proper definition of the error $\mathcal{E}$ should guarantee that all networks hybridized with the same threshold $\mathcal{E} < \mathcal{E}_t$ lead to the exact same evolution of $\tilde{h}_x$ and $\tilde{h}_y$ in time.

\begin{figure}[h!]
    \centering
    \includegraphics[width=0.9\textwidth]{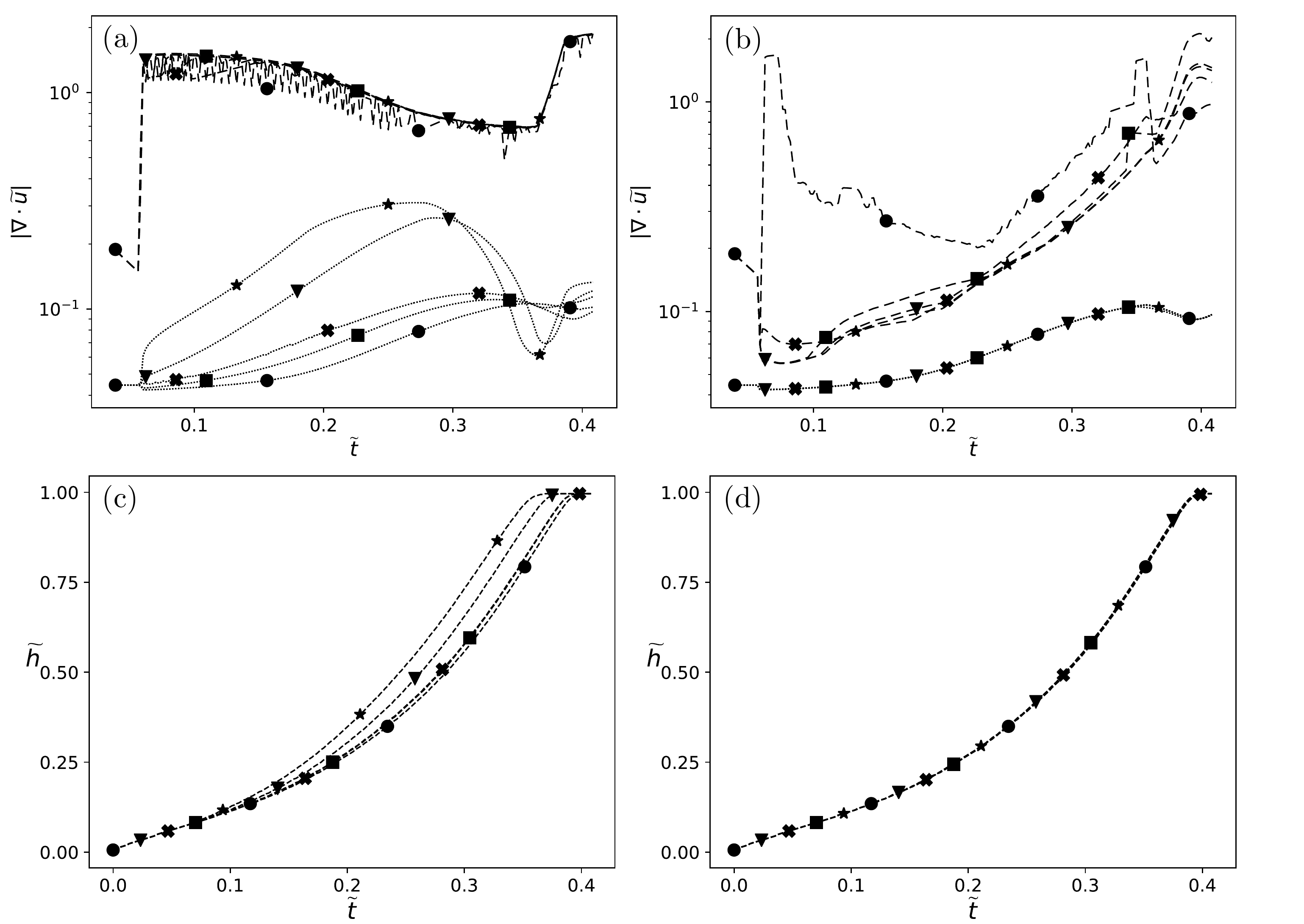}
    \caption{ $\mathcal{E}_{\infty}$ (\dashed) and $\mathcal{E}_{1}$ (\dotted), and plume head position $\tilde{h}_y$ (bottom), where (a, c) correspond to $\mathcal{E}_{t} = min(\mathcal{E}_{\infty})$ and (b, d)  $\mathcal{E}_{t} = min(\mathcal{E}_{1})$ evaluated on the no cylinder test case, at a $R_i=14.8$ obtained by several networks: $\blacktriangledown$ MonoScale, $\blacksquare$ MultiScale, $\times$ SmallScale, and $\bullet$ Unet, as well as the $\star$ Jacobi solver.}
    \label{fig:PH_max_mean}
\end{figure}

\begin{figure}[h!]
    \centering
    \includegraphics[width=0.9\textwidth]{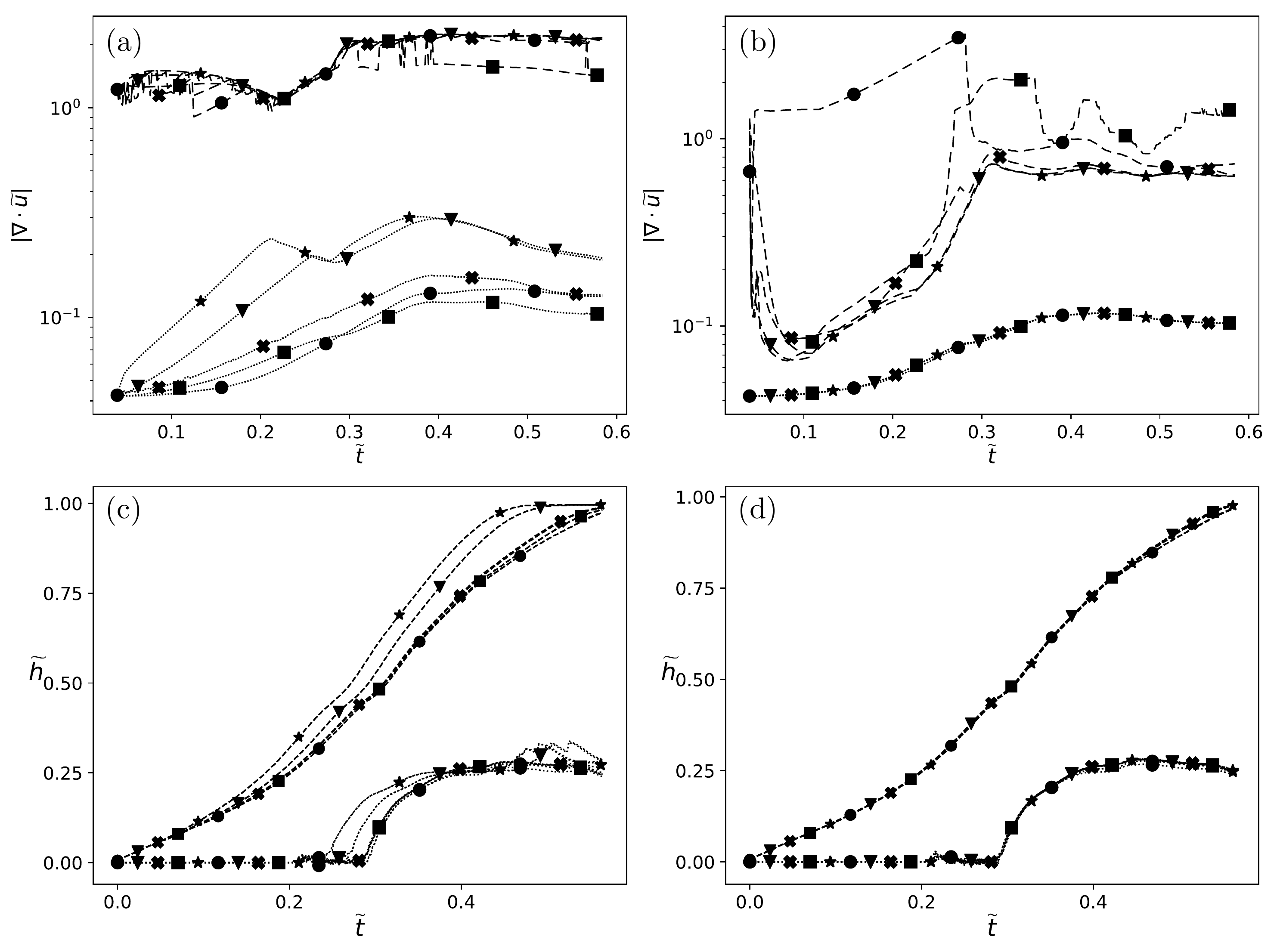}
    \caption{$\mathcal{E}_{\infty}$ (\dashed) and $\mathcal{E}_{1}$ (\dotted), and plume head position ($\tilde{h}_y$, $\tilde{h}_x$)  (bottom), where (a, c) correspond to $\mathcal{E}_{t} = min(\mathcal{E}_{\infty})$ and (b, d)  $\mathcal{E}_{t} = min(\mathcal{E}_{1})$ evaluated on the cylinder test case, at a $R_i=14.8$ obtained by several networks: $\blacktriangledown$ MonoScale, $\blacksquare$ MultiScale, $\times$ SmallScale, and $\bullet$ Unet, as well as the $\star$ Jacobi solver.}
    \label{fig:PH_max_mean_cyl}
\end{figure}

First, a test is carried out in order to choose between $\mathcal{E}_{1}$ and $\mathcal{E}_{\infty}$ as a proper indicator of the flow behavior, characterized by $\tilde{h}_x$ and $\tilde{h}_y$. The lowest values obtained in Fig.~\ref{fig:Basic_div} are chosen as threshold values. With these thresholds, the four different networks are evaluated and compared with the Jacobi method using a number of iterations also driven by the same threshold. Results on the divergence errors and head's position are displayed in Figs.~\ref{fig:PH_max_mean} and~\ref{fig:PH_max_mean_cyl} for the two test cases. First, it can be noticed that the specified thresholds are correctly followed by the networks and the Jacobi methods, since for the maximum error (Sub-figure-a in Figs.~\ref{fig:PH_max_mean} and~\ref{fig:PH_max_mean_cyl}) and mean error (Sub-figure-b in Figs.~\ref{fig:PH_max_mean} and~\ref{fig:PH_max_mean_cyl}), the divergence curves are superimposed. This demonstrates how the hybrid approach is able to guarantee a level of accuracy for the neural network predictions, which are otherwise unreliable. The resulting head positions $\tilde{h}_x$ and $\tilde{h}_y$, representative of the physical behavior of the simulations, are also provided (Sub-figure-c and Sub-figure-d in Figs.~\ref{fig:PH_max_mean} and~\ref{fig:PH_max_mean_cyl}). Interestingly, it is found that imposing the same mean error to all methods yields exactly the same time evolution of the physical targets, for both cases with and without obstacles. However, imposing the maximum error as a threshold leads to different, yet close, simulation behaviors. Typically, the same trend as in Fig.~\ref{fig:Basic_head} is observed, with a too rapid rise of the plume when predicted by the Jacobi method ($\star$) and the MonoScale network ($\blacktriangledown$). This faster convection speed is highlighted in the second case, where the plume is impinging the cylinder obstacle at shorter times for these two methods, resulting in an early flow deviation characterized by a non-null $\tilde{h}_x$ position (Fig.~\ref{fig:PH_max_mean_cyl}-c). Note however that prescribing a lower maximum error, typically $\mathcal{E}_t = 0.18$, yields also a similar plume evolution for all networks (\ref{sec:appendix_threshold}). %{\color{red} Pour info du coup, si tu augmentes le threshold pour l'erreur mean, les réseaux se comportent de la même façon ou non ? En gros, est-ce que le threshold base sur le mean est aussi sensible à la valeur de Et ou pas ? }. 

\subsection{Analysis of the error distribution}

\begin{figure}[h!]
    \centering
    \includegraphics[width=0.99\textwidth]{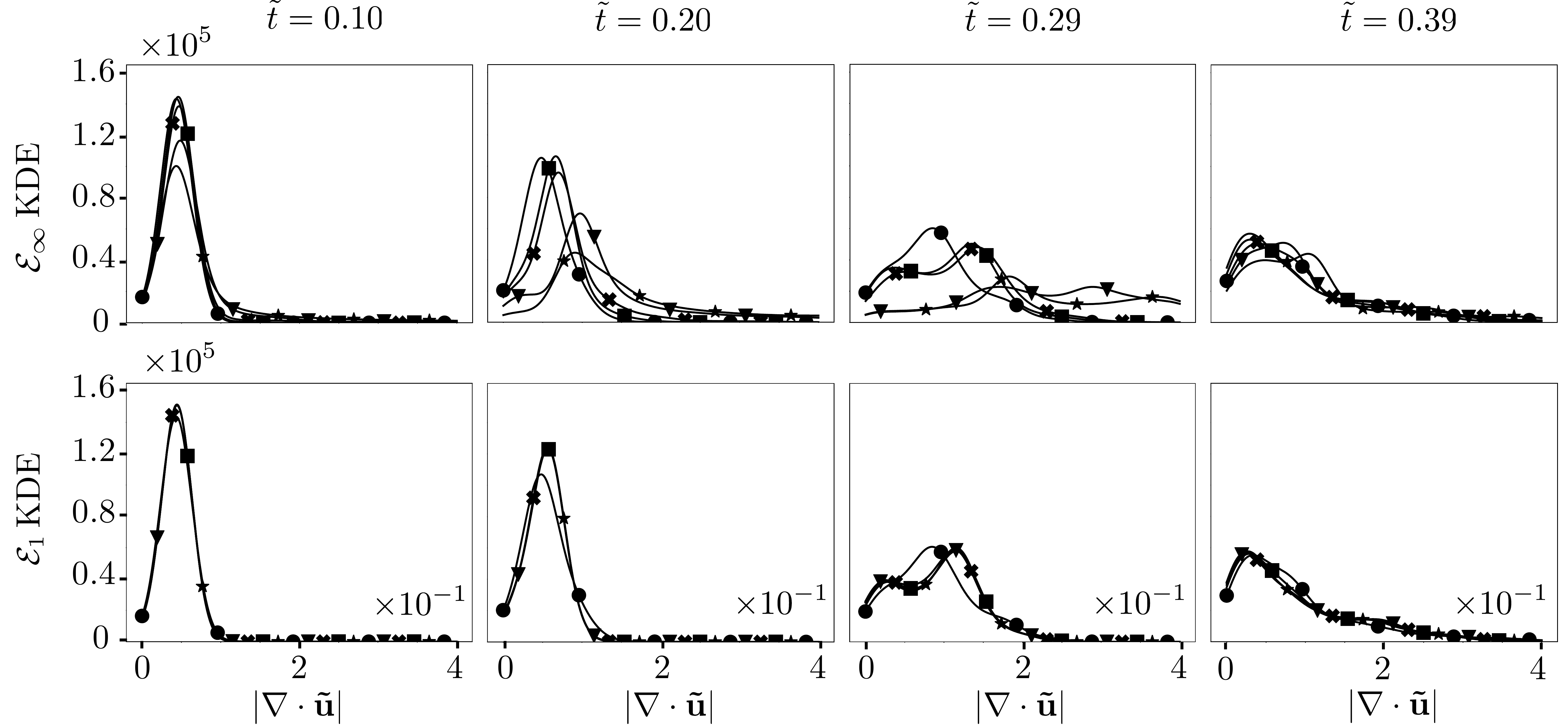}
    \caption{KDE at 4 times ($\tilde{t}$ = 0.10, 0.20, 0.29 and 0.39) of the cases where $\mathcal{E} = \mathcal{E}_{\infty}$ (top) and $\mathcal{E} = \mathcal{E}_{1}$ (bottom) of the no cylinder test case, at a $R_i=14.8$ obtained by several networks: $\blacktriangledown$ MonoScale, $\blacksquare$ MultiScale, $\times$ SmallScale, and $\bullet$ Unet, as well as the $\star$ Jacobi solver. %{\color{red} Il manque la legende dans la caption ... il faut que dans tes figures tous les textes soient de même taille, axe comme les t=xx}
    }
    \label{fig:KDE_mean_or_max_nocyl}
\end{figure}

To further understand how the error definition ($\mathcal{E}_{1}$ or $\mathcal{E}_{\infty}$) is driving the overall behavior of the simulations, the time-evolution of the spatial distribution of the error is studied. To do so, at a given normalized time $\tilde{t}$, the distribution of the divergence error is computed by a Kernel Density Estimation (KDE,~\cite{john2013estimating}). The evolution of the KDE for the plume case without obstacle is displayed in Fig.~\ref{fig:KDE_mean_or_max_nocyl}. The same analysis on the second test case corresponding to the plume impinging the cylinder is performed in~\ref{sec:appendix_cylinderKDE}. Results are displayed at four timesteps: $\tilde{t} = 0.1$, $0.2$, $0.29$ and $0.39$. Both $\mathcal{E}_{\infty}$ (top) and $\mathcal{E}_{1}$ (bottom) have been tested. The corresponding threshold values $\mathcal{E}_{t}$ have been set to the lowest values of the error obtained by all networks without hybrid approach (Section~\ref{sec:no_hybrid}). When following the maximum divergence threshold (top), it can be noticed that the error distributions differ depending on the method employed. A significant difference is observed at $\tilde{t} = 0.1$ for the Jacobi iterative method, for which the error distribution is spreading over a wide range, from $0$ to $0.3$. In comparison, all neural networks at this normalized time produce an error close to $0.05$, resulting in a sharp unimodal density function. Note that such behavior is ideal since the error is well controlled in space, yet the threshold value ($\mathcal{E}_t = 1.5$ at $\tilde{t} = 0.1$, Fig.~\ref{fig:PH_max_mean}-a) is not directly linked to the most probable error ($\mathcal{E} \approx 0.05$ at $\tilde{t} = 0.1$). At the same time $\tilde{t} = 0.1$, using $\mathcal{E}_{1}$ (Fig.~\ref{fig:KDE_mean_or_max_nocyl}, bottom) generates exactly the same error distribution for all approaches, including the Jacobi iterative method. This result generalizes the conclusions established from Figs.~\ref{fig:PH_max_mean} and~\ref{fig:PH_max_mean_cyl}: when specifying the error as the spatially averaged $\mathcal{E}_{1}$ error on the CFD domain, the hybrid approach yields exactly the same error distribution (whatever the method used or the network architecture), and consequently produces the same evolution of the plume head position $\tilde{h}$, i.e., the same physical behavior of the simulation. Interestingly, the error distribution is similar to the case following $\mathcal{E}_{\infty}$: the probability density function of the divergence error is unimodal, with a peak located at $|\nabla \cdot \tilde{u}| \approx 0.05$. However, compared with the use of the maximum error, here the threshold value ($\mathcal{E}_t \approx 0.05$ at $\tilde{t} = 0.1$, Fig.~\ref{fig:PH_max_mean}-b) directly corresponds to the most probable error.

As the simulation is running, a clear advantage of using $\mathcal{E}_{1}$ can be observed at $\tilde{t} = 0.2$. Not only it yields the same distribution error for all methods, but this distribution is close to the one obtained at previous times (e.g. $\tilde{t} = 0.1$), revealing that the hybrid approach is consistent in time. On the contrary, the use of $\mathcal{E}_{\infty}$ leads to very different error distributions. It is worth noting that the Jacobi method ($\star$) and the MonoScale ($\blacktriangledown$) exhibit a distribution shift towards higher error values, thus explaining their higher mean values (Fig.~\ref{fig:PH_max_mean}-a). This result is consistent with the unphysical behavior of these two simulations (too fast rising of the plume), since the evolutions of $\tilde{h}$ are strongly correlated with the mean value error (Figs.~\ref{fig:PH_max_mean}-b and d). 

At larger timesteps ($\tilde{t} = 0.29$ and $0.39$), the error distribution is spreading for all cases toward higher error levels (from $0$ to $0.3$ for most cases). Again, defining the error and threshold values using $\mathcal{E}_{\infty}$ (top) is not able to control the whole error distribution. The Unet ($\bullet$) produces the most unimodal distribution with a peak at a low error of $0.05$, while the Jacobi method ($\star$) and MonoScale network ($\blacktriangledown$) generate a flat error distribution with larger errors up to $0.6$. However, following $\mathcal{E}_{1}$ (bottom) is still producing consistent error distributions, with low discrepancies between the various methods, including the MonoScale and Jacobi. A slight shift of the error distribution can be observed, with a most probable error close to $0.1$ (instead of $0.05$ for early times). This is due to the fact that the present test case is performed with a varying threshold value $\mathcal{E}_t(\tilde t)$: this value is chosen as the lowest error value obtained without the hybrid approach, which is therefore not fixed in time. Figure~\ref{fig:PH_max_mean}-b reveals that the threshold value of the mean error is actually increasing in time, from $0.05$ to early timesteps towards $0.1$ at the end of the simulation ($\tilde{t} = 0.6$).

Overall, the description of the error $\mathcal{E}$ and threshold value $\mathcal{E}_t$ using the spatially averaged error over the CFD domain has shown a remarkable ability to control the whole distribution error in time and space, whatever the method used. For all times, the distribution is unimodal, with a most probable error in the range $0.05-0.1$, in good agreement with the threshold value of the mean error employed for the hybrid approach. This is not surprising since the error distribution is a unimodal symmetric probability function, for which the most probable and mean values are equal. In other words, prescribing the mean error is constraining the whole error distribution, whereas following the maximum value $\mathcal{E}_{\infty}$ only constrain the distribution tail, which seems insufficient to control the physics of the system, in particular here the rising speed of the plume head. Note that similar trends and conclusions are observed with the plume impinging the cylindrical obstacle (Fig.~\ref{fig:KDE_mean_or_max_cyl} in \ref{sec:appendix_cylinderKDE}), even if the error distributions are no more unimodal: the hybrid approach based on the mean error is still able to control the whole distribution error whatever the network architecture. For completeness, the whole spatial distributions of the error are displayed for the two cases (with and without obstacle) and both error definitions ($\mathcal{E}_{1}$ and $\mathcal{E}_{\infty}$) in \ref{appendix:heatmaps}. 

Finally, this study reveals that the neural networks hybridized with a Jacobi method are able to guarantee the accuracy level of the whole simulation in time, in particular when choosing an error defined as the spatially averaged divergence error in the CFD domain. Consequently, in the following, the hybrid approach based on $\mathcal{E}_{1}$ is chosen to ensure that all networks have a similar accuracy with the same plume head evolution in time, a pre-requisite to perform a fair comparison of the network performances. 

\section{Performance assessment}
\label{sec:performance}

In this section, the performance of neural networks is assessed for the time of inference, i.e., the time needed to produce the pressure correction. The analysis is innovative in two ways: (i) a fair performance comparison is performed using the hybrid approach, allowing the assessment of the time of inference at a fixed error level whatever the method (Jacobi or neural network architecture), and (ii) these performances are evaluated for several grid sizes. Note that even if CNN are theoretically capable of dealing with CFD domains of arbitrary size and resolution, they have in practice difficulties which such cases, probably because of boundary effects~\cite{kayhan2020translation}: here again, the hybrid approach is useful to guarantee the error level of the solution when changing the grid size. To do so, as suggested in Section~\ref{sec:control_error}, the mean divergence error is used to define the threshold $\mathcal{E}_t$. All simulations were performed on the same GPU card, namely here a NVIDIA Tesla V100 with 32 Gb of memory. Note that the performance evaluation on multiple GPUs is out of the scope of the present study. 

To measure the fluid solver performance, the following times are defined:

\begin{itemize}
    \item \textbf{$t_{inf}$}: Inference time taken by a neural network or the Jacobi solver to output one pressure field when the divergence field is inputted, i.e., not taking into account the time to correct the velocity field or the time spent on the extra Jacobi iterations of the hybrid solver. 
    \item \textbf{$t_{p}$}: Time taken to perform the entire pressure projection step. This includes the time to perform the first pressure inference, the extra Jacobi iterations in the hybrid process, as well as the correction of the velocity field.
    \item \textbf{$t_{it}$}: Time taken to perform an entire iteration of the fluid solver, i.e., the advection and pressure projection steps. 
\end{itemize}

\subsection{Network Comparison}

Using the hybrid strategy, a non-trivial trade-off has to be made between accuracy (usually implying more complex and deeper architectures) and fast inference time (requiring small networks). In practice, a less accurate network with a fast inference time will require more Jacobi iterations to reach the desired accuracy level, thus limiting drastically its performance. Comparing the performance of the several network architectures used in this work will provide a first insight on this trade-off, in order to establish guidelines for future developments of AI-based solvers.

\begin{table}[!t]
\label{table:network_inference}
\caption{Inference time of each network to produce the pressure field without Jacobi iterations, as well as the inference time of a Jacobi solver performing a single iteration}
\centering
\begin{tabular}{|c|c|c|c|c|c|}
\hline
\textbf{Time}          & \textbf{MonoScale} & \textbf{MultiScale} & \textbf{Unet} & \textbf{SmallScale} & \textbf{Jacobi (1 it)} \\ \hline
$t_{inf}$ (ms)         & 14.6      & 11.3       & 5.25 & 5.35 & 2.31      \\
$t_{inf}$/$t_{it}$ (\%) & 29.1      & 24.1       & 12.9 & 13.1 &   6.10     \\ \hline
\end{tabular}
\end{table}

\begin{figure}[h!]
    \centering
    \includegraphics[width =0.95\textwidth]{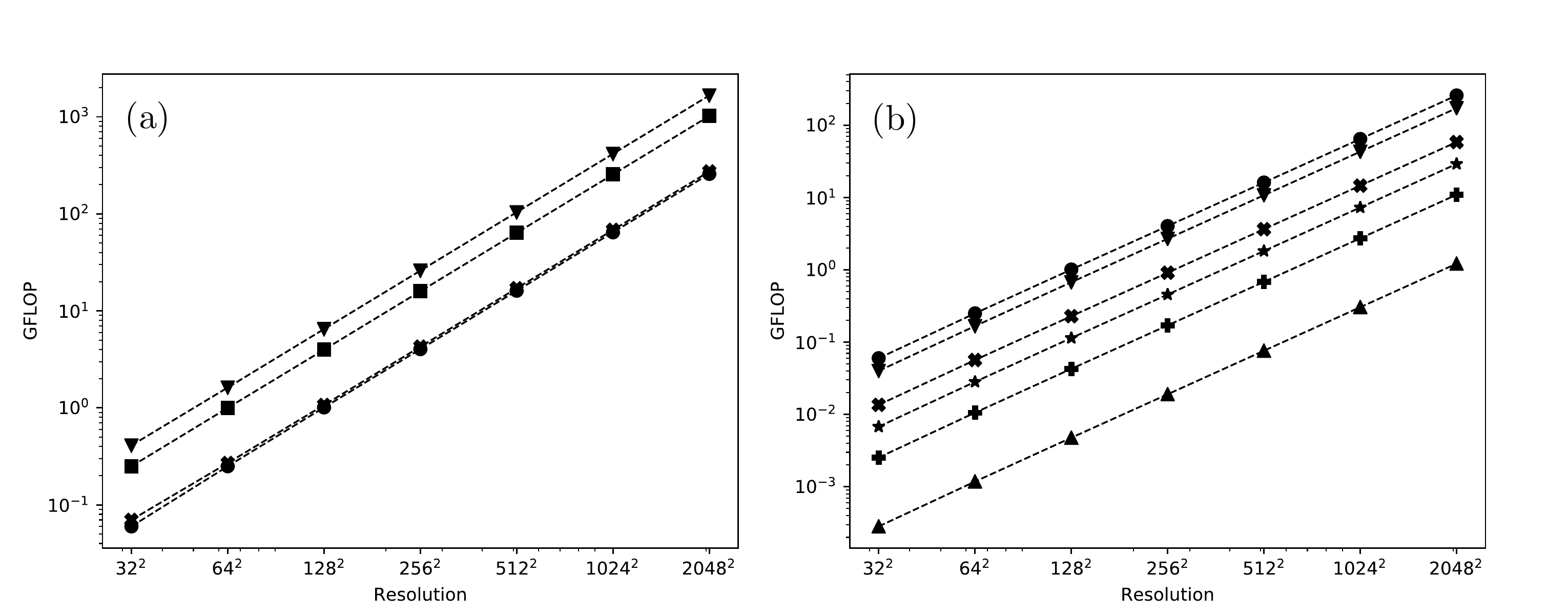}
    \caption{Evolution of the number of Floating-point operations (in Giga units) with the domain size (varying from 1024 to 4.2~$10^{6}$ cells) needed in a single network inference for the four studied networks (a): $\blacksquare$ MultiScale, $\times$ SmallScale, $\blacktriangledown$ MonoScale and $\bullet$ Unet, and for the scales composing the Unet network (b):$\bullet$ Unet, $\blacktriangledown$ $n^2$, $\blacktriangledown$ $n_{1/2}^2$,  $\times$ $n_{1/2}^2$, $\star$ $n_{1/4}^2$, + $n_{1/8}^2$ and $\blacktriangle$ $n_{1/16}^2$.}
    \label{fig:flops}
\end{figure}

Consequently, the performance assessment of the neural networks is split in two steps: (i) assess the inference time $t_{inf}$ of each CNN to produce the output, and (ii) evaluate the number of Jacobi iterations needed by each network, using the CNN prediction as initial guess, to reach the target accuracy. Table~\ref{table:network_inference} shows the inference time of each network, and reveals that $t_{inf}$ is not directly associated with the number of parameters of the network. Indeed, whereas MonoScale, MultiScale and Unet have the same number of parameters, they also have very different inference times. Similarly, the SmallScale contains three times fewer parameters compared with Unet, but still requires more time to output its prediction. This is due to the specific architectures of the Unet and MultiScale networks which involve several scales: filters and their associated parameters at low resolution (capturing the large scales of the flow field) are applied on reduced inputs, therefore limiting the number of operations to perform, and thus reducing $t_{inf}$ (Fig.~\ref{fig:flops}-b). Thus, the more parameters at lower scale, the faster the network is. This is confirmed by the associated number of floating-point operations (FLOP) for each network architecture, displayed in Fig.~\ref{fig:flops} for each network and several grid sizes, from $1024$ to $4.2 \times 10^6$ cells. It reveals that the FLOP number depends linearly on the grid resolution and is directly correlated with the inference times of Tab.~1: the $5$ scales of the Unet result in a low FLOP number, and therefore a reduced $t_{inf}$, comparable to the SmallScale network with three times fewer parameters. The MultiScale network also contains $3$ scales, but most parameters are concentrated at the larger resolution, which is therefore dominating the FLOP number and inference time. Only a small gain is observed for this architecture compared with a standard MonoScale network. 

\begin{figure}[h!]
    \centering
    \includegraphics[width=0.95\textwidth]{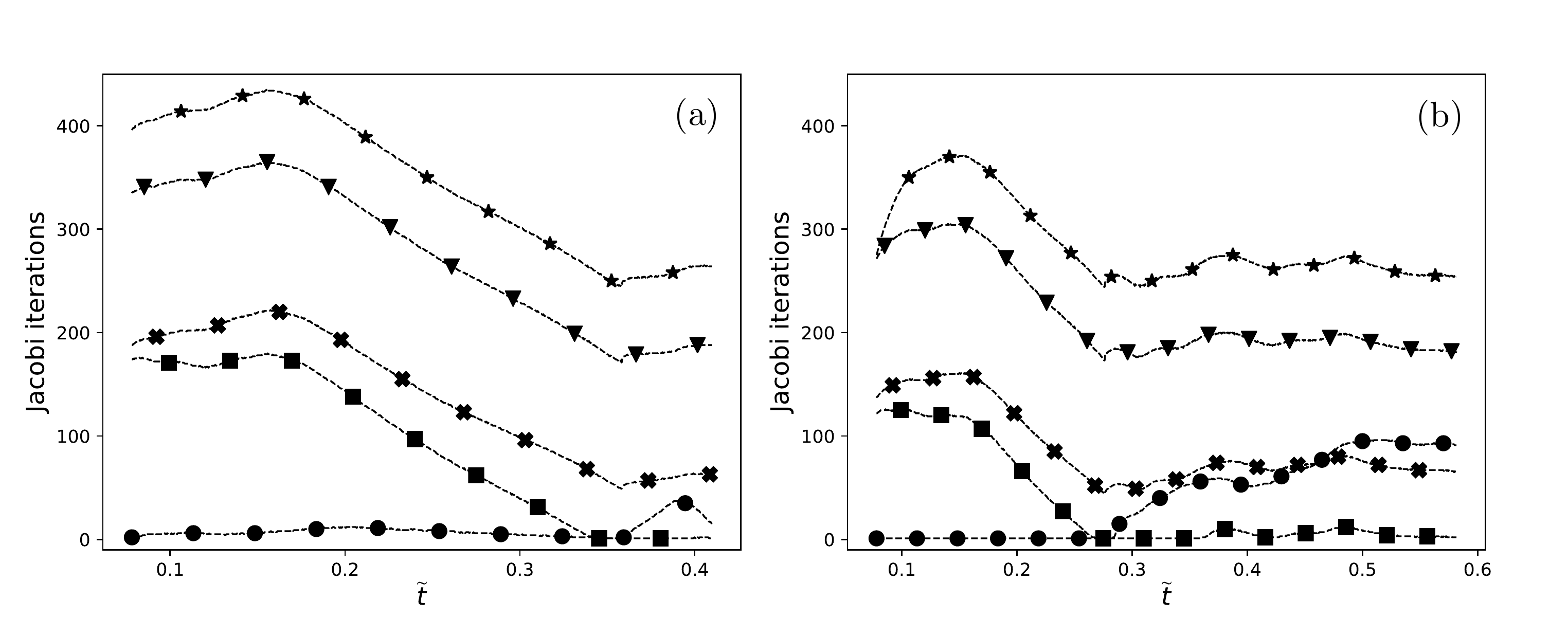}
    \caption{Number of Jacobi iterations needed to ensure $\mathcal{E}_t = min(\mathcal{E}_1)$ on the non-cylinder (a) and the cylinder (b) test cases with various networks: $\blacktriangledown$ MonoScale, $\blacksquare$ MultiScale, $\times$ SmallScale, and $\bullet$ Unet, as well as the $\star$ Jacobi solver. %{\color{red} faudrait peut etre ajouter en plus du nombre d'ite, un 2e axe des ordonnees à droite des figures ou on met le temps equivalent pour realiser ces iterations de Jacobi, afin d'avoir une idee de comparaison deja avec le temps du CNN seul, tableau 1}}
    }
    \label{fig:jacobi_its}
\end{figure}

Since the several networks have different error levels, the number of Jacobi iterations to reach the desired accuracy level is decisive in the overall performance of the AI-based code. Figure~\ref{fig:jacobi_its} shows the number of Jacobi iterations needed to reach the threshold $\mathcal{E}_t$, for both cases without (Sub-figure-a) and with (Sub-figure-b) obstacle. As expected, the classical Jacobi solver ($\star$) and the MonoScale network ($\blacktriangledown$) need numerous iterations, typically between $300$ and $400$. Using a multiscale architecture (either MultiScale or SmallScale) improves significantly this situation, since only $100$ to $200$ iterations are needed for the same accuracy level. Note that for the case with obstacle, the MultiScale network needs almost no Jacobi iteration after the plume has impinged the cylindrical obstacle ($\tilde{t} > 0.3$). In contrast, the Unet outperforms all networks, with almost no Jacobi iteration for all cases. However, it requires $50-100$ iterations after the plume has impinged the cylinder ($\tilde{t} > 0.3$), which is consistent with the previous finding that Unet has difficulties close to walls (Fig.~\ref{fig:heat_map_snapshots}). As a general conclusion, all networks outperform the classical Jacobi solver, proving that using neural networks as an initial guess to the iterative solver is relevant. However, this AI-based computational strategy is efficient only if the time needed to produce the initial guess by the CNN leads to a total time lower than using the Jacobi solver alone. 

\begin{figure}[h!]
    \centering
    \includegraphics[width=0.95\textwidth]{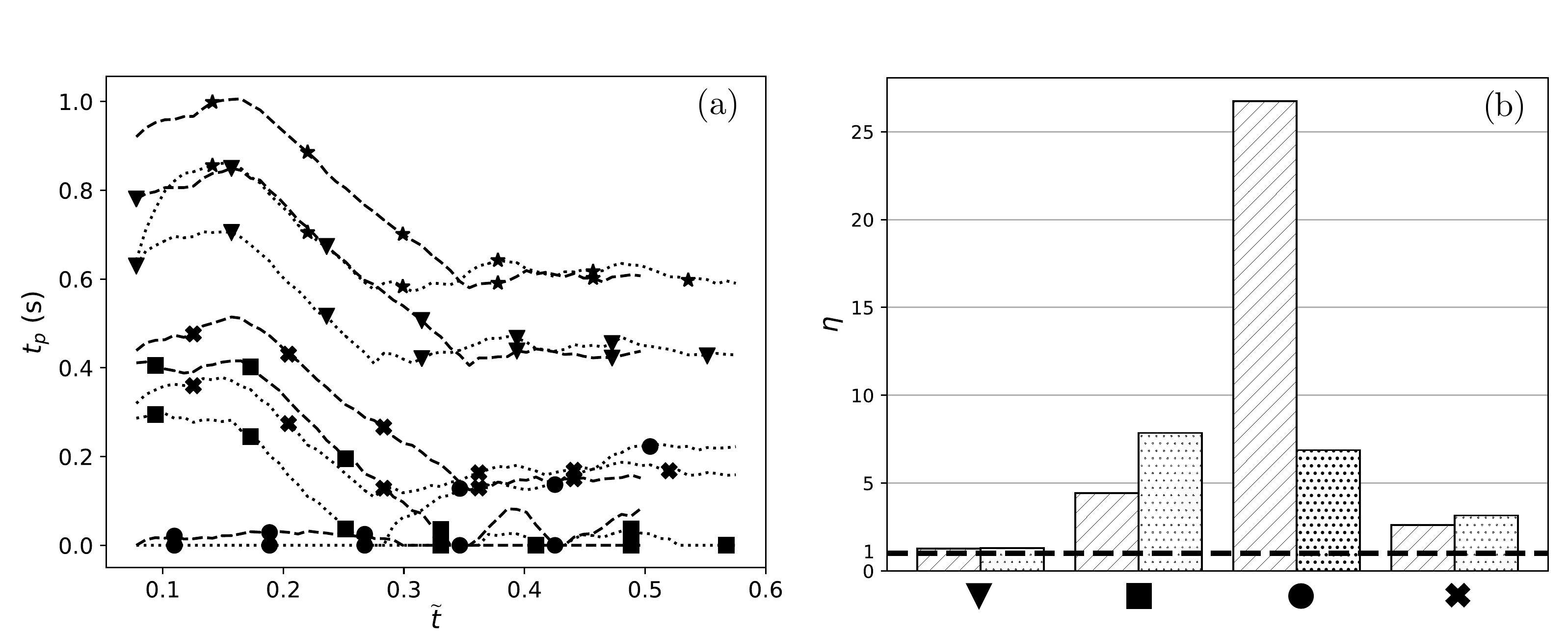}
    \caption{Time $t_p$ (a) and acceleration factor $\eta$ (b) for the non-cylinder test case (\dashed) and cylinder test case (\dotted), for the four studied networks: $\blacktriangledown$ MonoScale, $\blacksquare$ MultiScale, $\times$ SmallScale, and $\bullet$ Unet, as well as the $\star$ Jacobi solver. %{\color{red} ca me parait plus pertinent de mettre les barres en fonction du Jacobi comme reference, plutot que le MultiScale. Ainsi on aura directement le facteur d'acceleration des differentes methods, donc en gros plotte Tjacobi/Tnetwork. Ici c'est quoi l'iteration, juste Poisson ou Poisson+advection ?}}
    }
    \label{fig:time_mean}
\end{figure}

To verify this, the time to solve the Poisson equation ($t_{p}$) is now compared for all methods in Fig.~\ref{fig:time_mean}-a. As a comparison, the computational time to solve the advection equation is $t_{ad} = 35.5 ms$. The bar plot (Fig.~\ref{fig:time_mean}-b) highlights the acceleration factor $\eta = t_{p}^{jacobi} / t_{p}^{network}$ for each network architecture, where $t_{p}^{jacobi}$ and $t_{p}^{network}$ are the computational times $t_{p}$ of the Jacobi method and the hybridized network respectively.  First, it can be noticed that all AI-based approaches are faster than the Jacobi solver alone, which proves the efficiency of the proposed computational strategy using CNN to predict the solution of the Poisson equation. As expected, the MonoScale only provides a small gain, about $1.2-1.3$, since (i) it needs numerous iterations of the Jacobi solver to achieve the desired accuracy level, and (ii) it has the largest inference time (Tab.~1). Note that the latter is evaluated at $14.6ms$, which indicates that even for this network with large inference time, it is still negligible compared to the computational time of Jacobi iterations. Interestingly, the large (MultiScale) and small (SmallScale) multiscale networks achieve a comparable acceleration factor, respectively $\eta \approx 3.7$ and $2.4$ without obstacle, but differ when the plume impinges the cylinder, with $\eta \approx 7.5$ and $\eta \approx 3.1$. Indeed, the larger network achieves a better accuracy, thus limiting the number of Jacobi iterations. However, the accuracy-inference time trade-off is not trivial when dealing with deep neural networks embedded in a CFD solver, as in test cases where few Jacobi iterations are needed, the SmallScale could outperform the larger MultiScale network, thanks to the smaller network inference time. Finally, the Unet architecture outperforms all other methods in the case without obstacle with an acceleration factor $\eta = 26.1$. As Unet has difficulties handling boundaries, its performance drops in the case with obstacle, yet still accelerating the CFD solver by a factor $\eta = 6.6$, slightly below the MultiScale network performance. Note however that for longer simulations, the MultiScale network would further outperform Unet since requiring no Jacobi iteration for times larger than $\tilde{t} = 0.3$.

%{\color{red} Il faudrait finir avec un bref mot sur le temps complet diteration en prenant en compte l'advection, pour dire que maintenant, avec le UNet ou MultiScale, resoudre Poisson n'est plus forcement le bottleneck, car on passe de 80 pourcent du temps a 20 pourcent j'imagine... mettre les vrais valeurs bien sur.}

%\begin{figure}[h!]
%    \centering
%    \includegraphics[width=0.95\textwidth]{images/Hybrid/Plume/Ri_10/Div/Follow/time/Time_cyl.pdf}
%    \caption{Time taken to perform the pressure projection step per time step to ensure the $\mathcal{E}_t$ threshold (left) and the acceleration factor $\eta$ (right) for the cylinder test case, for the four studied networks: $\blacktriangledown$ MonoScale, $\blacksquare$ MultiScale, $\times$ SmallScale, and $\bullet$ Unet, as well as the $\star$ Jacobi solver. %{\color{red} tu es sur, le temps est bien en seconde la dans ces 2 figures ? 2s ou 6s ca me parait enorme, et pas coherent avec la figure qui suit quand je regarde ton graph a la resolution 512x512.}
%    }
%    \label{fig:time_cyl_mean}
%\end{figure}

\subsection{Performance evaluation on various grid sizes}

\begin{figure}[h!]
    \centering
    \includegraphics[width = 0.5\textwidth]{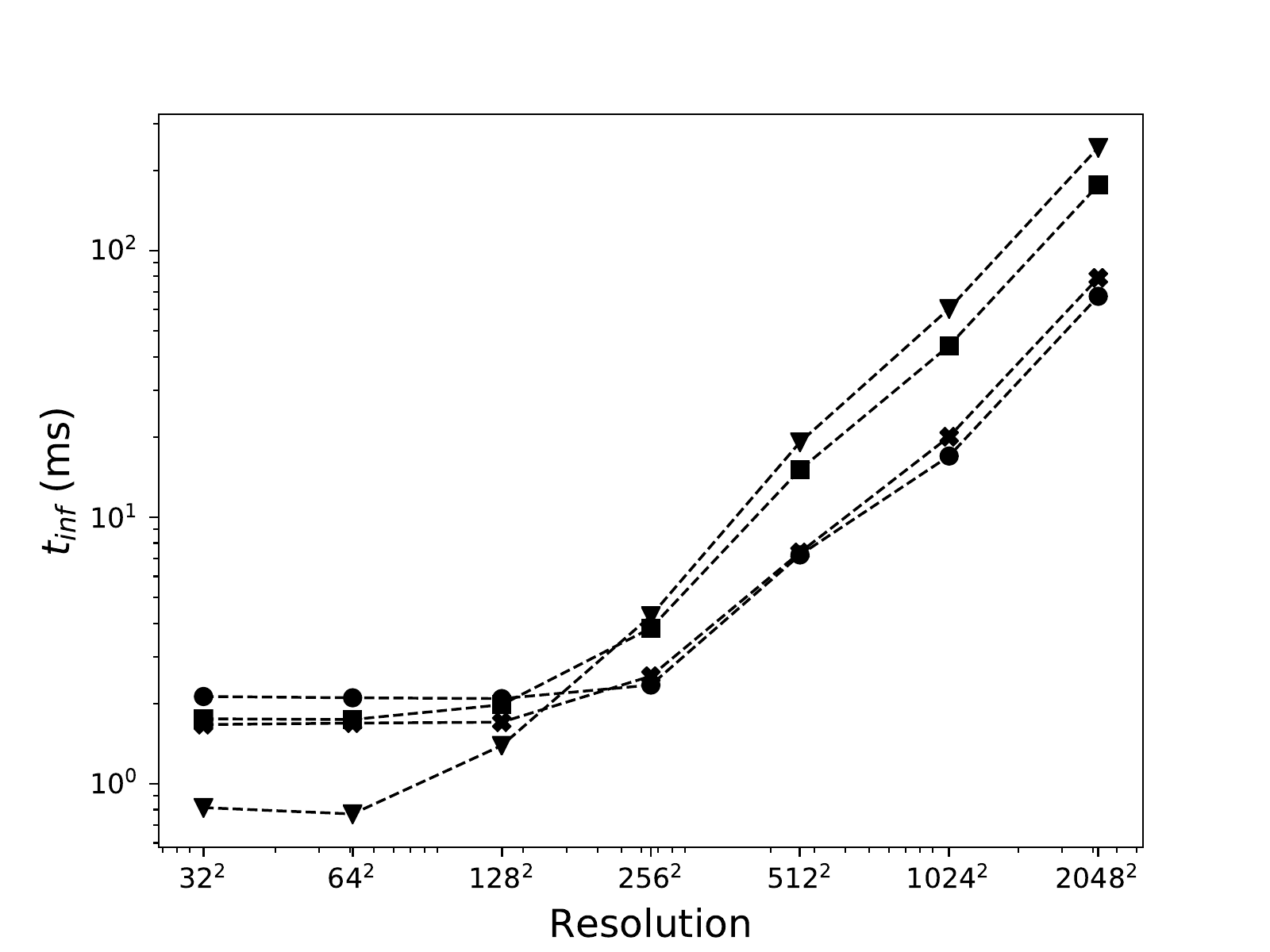}
    \caption{Time taken to perform the network prediction $t_{inf}$ for the four studied networks: $\blacktriangledown$ MonoScale, $\blacksquare$ MultiScale, $\times$ SmallScale, and $\bullet$ Unet, as well as the $\star$ Jacobi solver, on a grid size varying from 1024 to 4.2~$10^6$.}
    \label{fig:log_log_network_evolution_plt}
\end{figure}

The previous performance assessment was made at a fixed resolution of $2.6~10^5$ cells. Recent work~\cite{kayhan2020translation} revealed that networks can encode spatial location and resolution on the training dataset, which can lead to poor performance when applied to a different resolution than the one used during the training phase ($2.6~10^5$ cells in this work). This drawback is here tackled by the hybrid approach, which ensures the desired accuracy level whatever the grid size. It is also found in this work that the present CNN architectures are capable of generalizing to different grid sizes, even if no clear explanation can be given for this specificity. As a reminder, Fig.~\ref{fig:flops} showed that the number of operations per network evolves linearly with the grid size. However, Fig.~\ref{fig:log_log_network_evolution_plt} shows a different behavior when looking at $t_{inf}$, with two distinct regimes: (i) the network performances follow the same behavior as in Fig.~\ref{fig:flops} when the domain size is large enough, with a linear increase with grid size (ii) $t_{inf}$ remains constant for small grids. This phenomenon is related to the CPU overhead, as the time taken by the GPU kernels is inferior to the \textit{command transmission} between the CPU and the GPU, which depends only on the network architecture but not the grid size. Therefore, networks with a simpler architecture result in a lower CPU overhead, and so a faster inference. It should be noted how for example the MultiScale and SmallScale networks show a similar inference time when the GPU kernels are saturated, as the architecture and layer size are exactly the same. This saturation should be taken into account for small resolution test cases, for instance, by designing dedicated simple architectures.

%{\color{red} tu monres que le CNN la, ou la prediction totale CNN+iteration de Jacobi? Si c'est que le CNN, il faudrait ajouter a côte une figure avec le temps complet pour Poisson afin de voir l'effet de la resolution sur le facteur d'acceleration eta}

\subsection{Analysis of the network performances}

\begin{figure}[h!]
    \centering
    \includegraphics[width =0.99\textwidth]{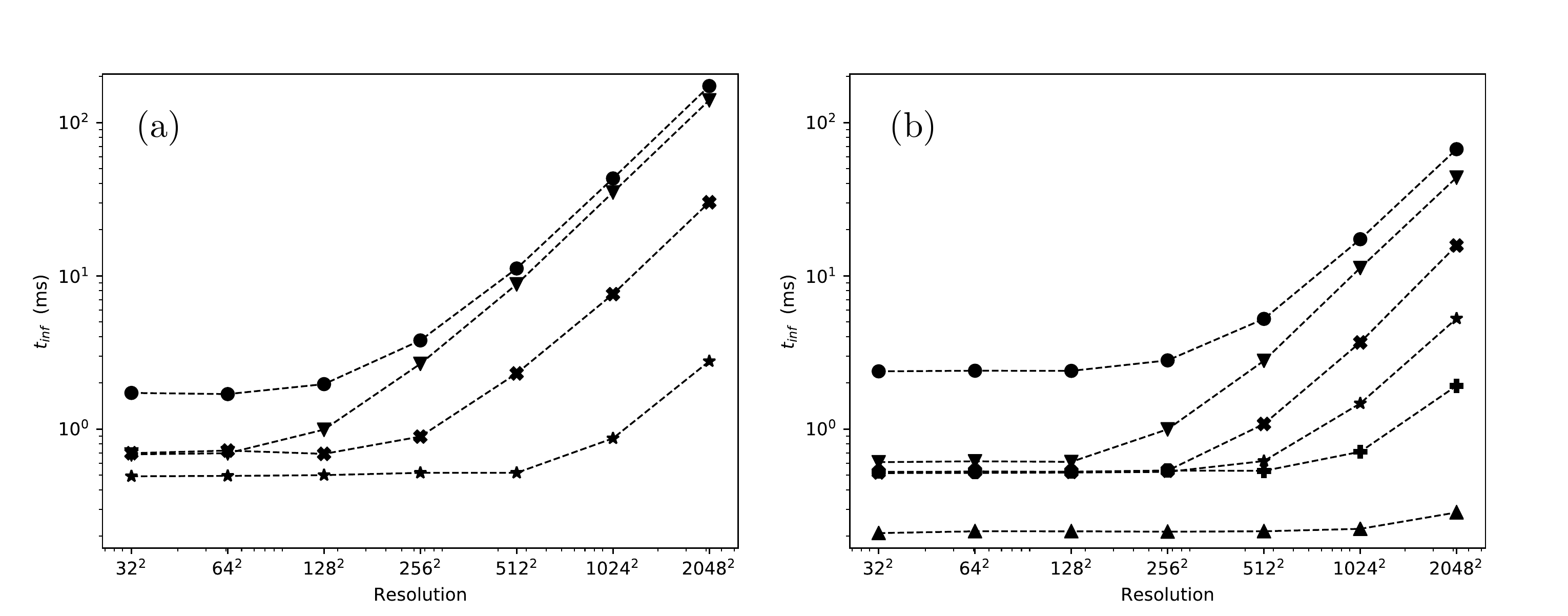}
    \caption{Time taken by each scale ($\bullet$ $n^2$, $\blacktriangledown$ $n_{1/2}^2$,  $\times$ $n_{1/2}^2$, $\star$ $n_{1/4}^2$, + $n_{1/8}^2$ and $\blacktriangle$ $n_{1/16}^2$) for the MultiScale (a) and Unet (b) networks, to perform a single inference on a grid size varying from 1024 to 4.2~$10^{6}$ cells. %  {\color{red} Enleve N+4+N/2 ce n'est pas utile je pense. Par contre tu pourrais ajouter la meme figure a droite avec les differentes echelles du Unet, avec le meme range de temps que Multiscale pour bien voir pourquoi le UNet a de meilleur perfo en temps}
    }
    \label{fig:log_log_resolution_evolution_scales_plt}
\end{figure}

In order to better understand the link between the network architecture and its performances, the time taken for different scales is displayed in Fig.~\ref{fig:log_log_resolution_evolution_scales_plt}, for both the MultiScale (a) and Unet (b) networks when the spatial resolution is varying. This figure shows the evolution of the 3 main scales of the MultiScale network, depending on the spatial resolution. As expected, the largest scale takes the major part of the inference time for both networks, performing most of the operations due to the larger size of the domain. Moreover, it is shown that the smaller scales are more prone to the command overhead, which makes the saturated regime occurring on a larger resolution range, for instance, the scale n/4 of the MultiScale is saturated until the resolution $512x512$, and the scale n/16 of the Unet until $1024x1024$. Exploiting smaller scales with more parameters could open the path to new computational strategies with large computational domains divided into multiple smaller subdomains of the appropriate size, for example, adapted to each scale of the network.

\section{Conclusions}

This work focuses on the resolution of the Poisson equation with convolutional neural networks applied to incompressible fluid flows. A plume test case, with and without obstacle, is chosen since parametrized by a single dimensionless quantity: the Richardson number. The well-known MultiScale and Unet architectures are analysed, and compared to a simpler architecture as well as with a traditional Jacobi solver. In order to ensure a user-defined accuracy level, a hybrid strategy is used where the neural networks are coupled with a Jacobi solver when a threshold error criterion is not satisfied. This work shows that a threshold based on the mean divergence of the flow field ensures a consistent physical behavior of the fluid flow. The various networks are compared in both accuracy and inference time when using the hybrid strategy. It is found that the network behavior varies with the studied case. When there is no obstacle in the domain, the Unet network outperforms the other networks both in accuracy and performance. However, when an obstacle is introduced, its accuracy decreases because of difficulties close to walls, resulting in a simulation time superior to the MultiScale network. For both cases, the MultiScale and Unet provide speed-ups of order 5-7, demonstrating the potential of such a method. Moreover, it is shown that the multiple scales of the Unet and MultiScale networks result in a faster inference time, as long as the resolution is larger that $256x256$ . For lower resolutions, the CPU overhead controls the inference time, enabling the simpler MonoScale network to obtain the fastest inference time. As a conclusion, this work has revealed the potential of coupling CNN with multiple scales to a CFD solver to produce fast and reliable flow predictions.

% while maintaining the desired accuracy level.

\section{Acknowledgement}
%{\color{red} put here CALMIP, maybe Jolibrain as well and maybe Munich}

This work was founded by the CERFACS institute and ISAE-SUPAERO in Toulouse, France. The calculations were performed using HPC resources from CALMIP on Olympe (Grant 2020-p20035). The presented fluid solver was initially developed by Antonio Alguacil in collaboration with the Jolibrain company, who also offered their technical support throughout this project. Finally, the Physics-based Simulation group (Thuerey group) of the TUM (Technical University of Munich) should be acknowledged for their support.  

%Bibliography
\bibliographystyle{unsrt}  
\bibliography{references}  
\clearpage
\appendix

\clearpage
\section{Spatial distribution of the divergence error}
\label{appendix:heatmaps}

To complement this study in Section~\ref{sec:errorquant}, the divergence distribution at four timesteps is shown here. Note that these divergence percentiles show the spatial distribution of the divergence, but they do not give any information about the absolute value of the studied error.

\begin{figure}[h!]
    \centering
    \includegraphics[width =1.0\textwidth]{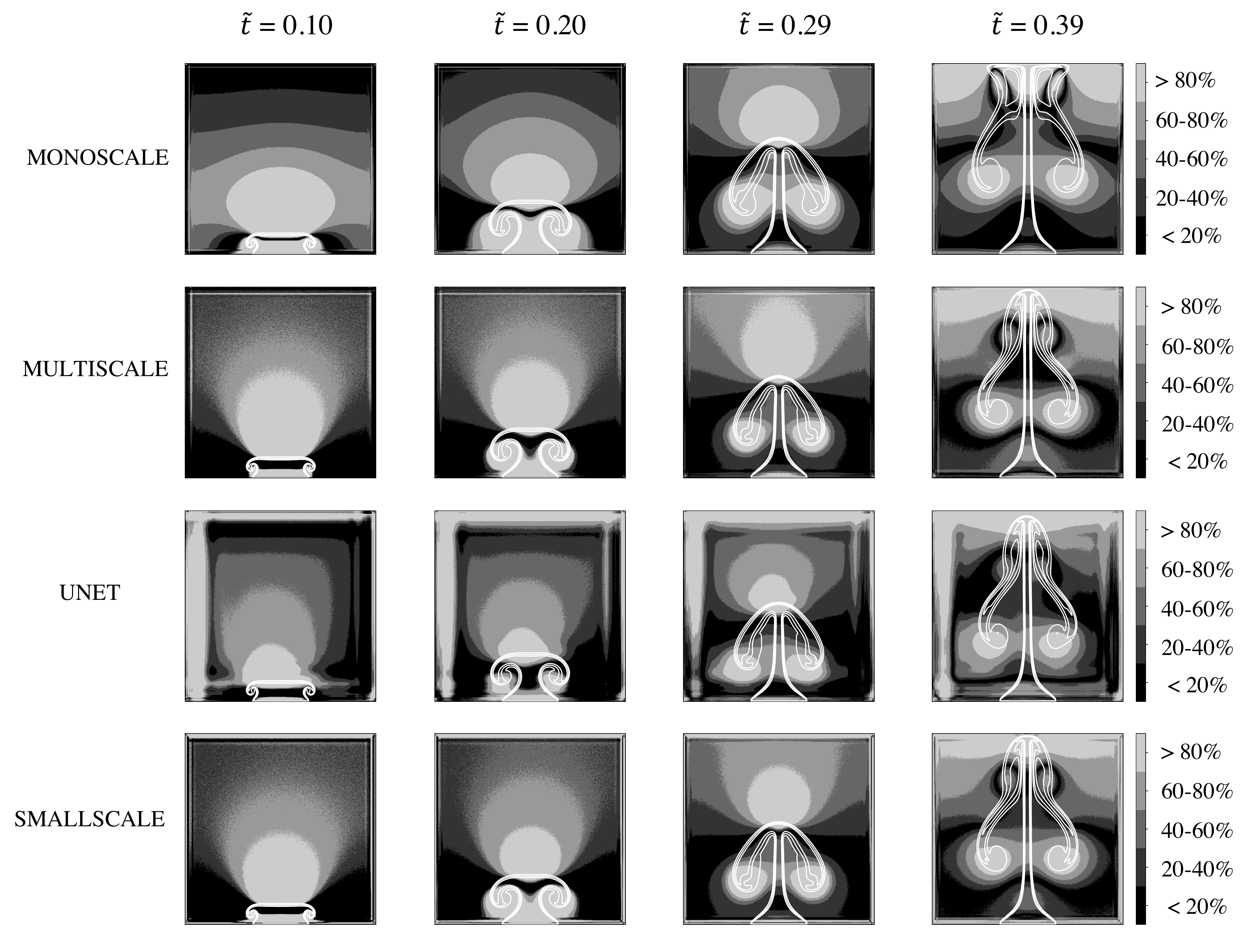}
    \caption{Divergence percentiles and density iso-contours (in white) of the four studied networks for the no cylinder case at timesteps $\tilde{t}$~=~0.10, 0.20, 0.29 and 0.39 when the networks prediction are not coupled with the hybrid strategy. }
    \label{fig:initial_field_nocyl}
\end{figure}

Figure~\ref{fig:initial_field_nocyl} represents the divergence distributions at 4 timesteps for the no cylinder case where no hybrid strategy is used, i. e., the divergence distribution of the raw output of the four networks. Qualitatively, all networks behave similarly, as the main divergence sources are found on the plume head and vortices. However, when closely analysed, each network counts with its distinct behavior related to the network architecture. The MultiScale and SmallScale architectures behave similarly, showing a smooth divergence distribution, with the presence of a divergence stride close to the domain edges, probably related to the network padding, while the Unet network seems to struggle predicting the pressure field close to the walls (particularly the top and left wall). Moreover, the MonoScale network shows the smoothest divergence distribution, even if its flow topology clearly shows a faster plume propagation.

\begin{figure}[h!]
    \centering
    \includegraphics[width =1.0\textwidth]{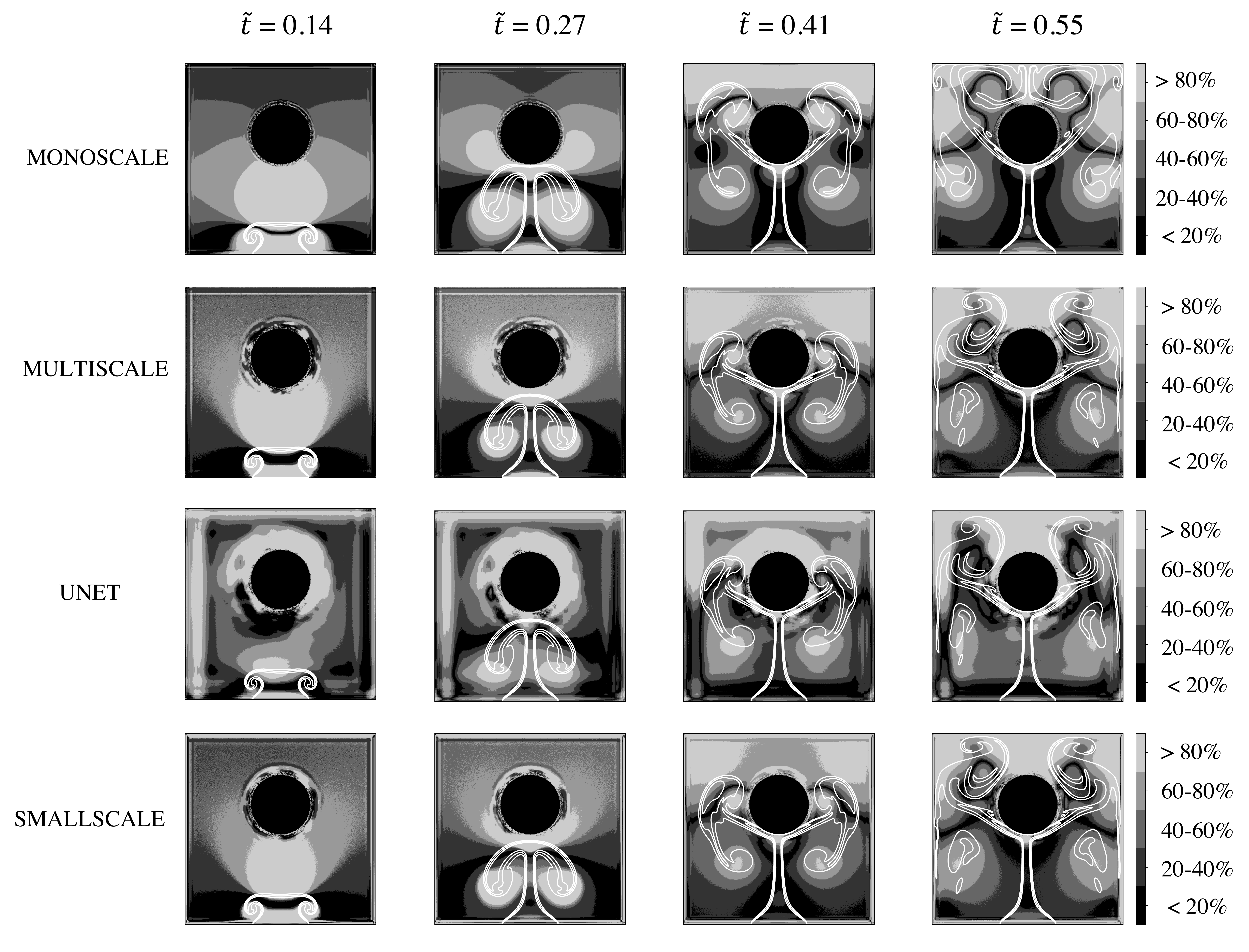}
    \caption{Divergence percentiles and density iso-contours (in white) of the four studied networks for the cylinder case at timesteps $\tilde{t}$~=~0.14, 0.27, 0.41 and 0.55 when the networks prediction are not coupled with the hybrid strategy.}
    \label{fig:initial_field_cyl}
\end{figure}

Similarly, Fig~\ref{fig:initial_field_cyl} represents the same divergence distribution but on the cylinder case. The divergence distribution follows a similar behavior as the one observed in Fig.~\ref{fig:initial_field_nocyl}. However, the presence of the cylinder highlights potential difficulties of networks to predict flow-wall interactions. In particular, it is shown that the Unet network particularly faces such a difficulty, as the major part of its error concentrates near the cylinder wall. Analysing the flow topology, shows that the MonoScale network creates larger structures, probably due to an incorrect prediction of the baroclinic-torque.

\begin{figure}[h!]
    \centering
    \includegraphics[width=0.99\textwidth]{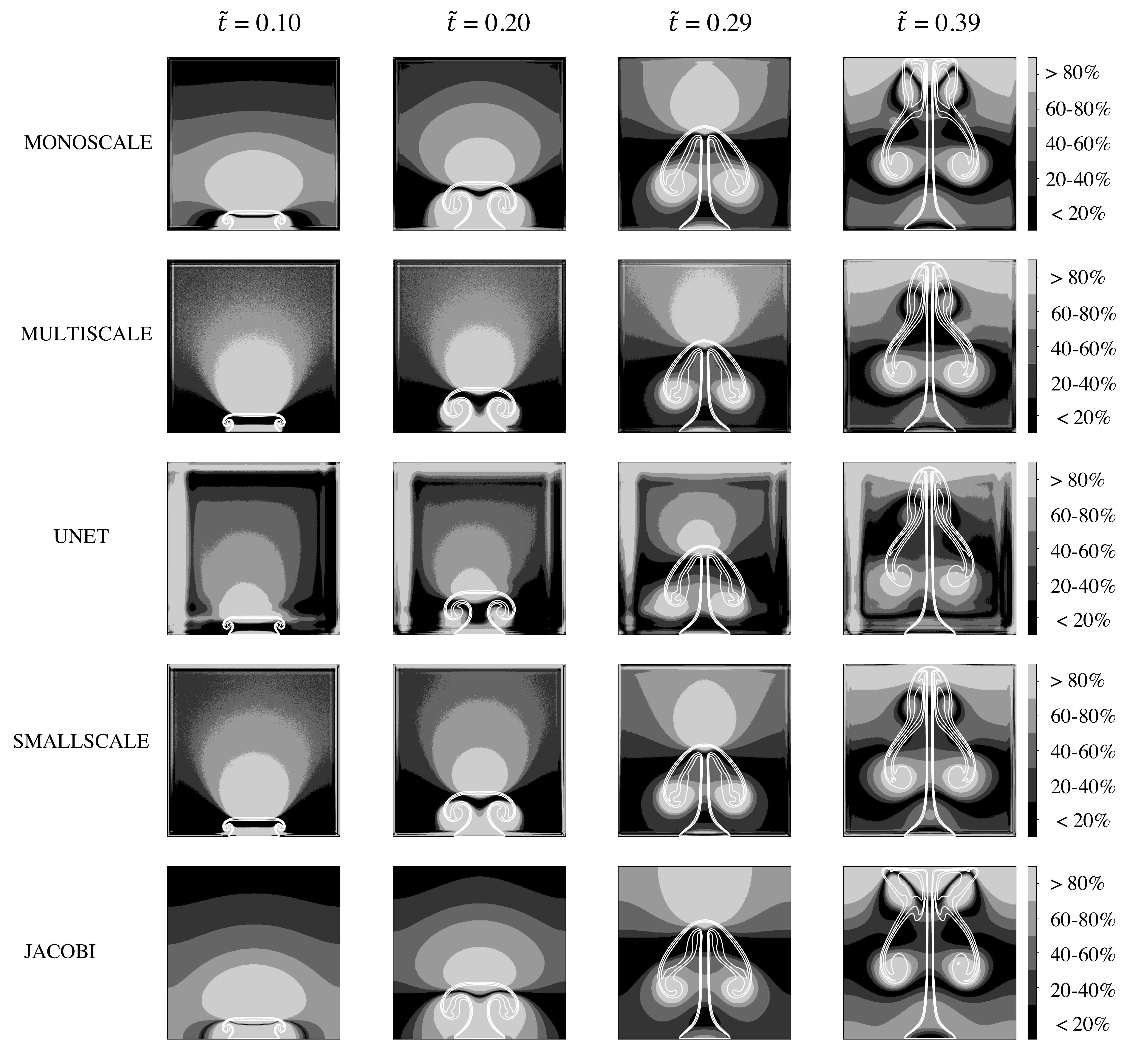}
    \caption{Divergence percentiles and density iso-contours (in white) of the four studied networks for the no cylinder case at timesteps $\tilde{t}$~=~0.10, 0.20, 0.29 and 0.39 when the networks prediction are coupled with the hybrid strategy. The chosen threshold is $\mathcal{E} = \mathcal{E}_{\infty}$, where $\mathcal{E}_{t} =  min(\mathcal{E}_{\infty})$ obtained by any network at a given timestep, without being coupled with the hybrid strategy.}
    \label{fig:heat_map_max_div_no_cylinder}
\end{figure}

\begin{figure}[h!]
    \centering
    \includegraphics[width=0.99\textwidth]{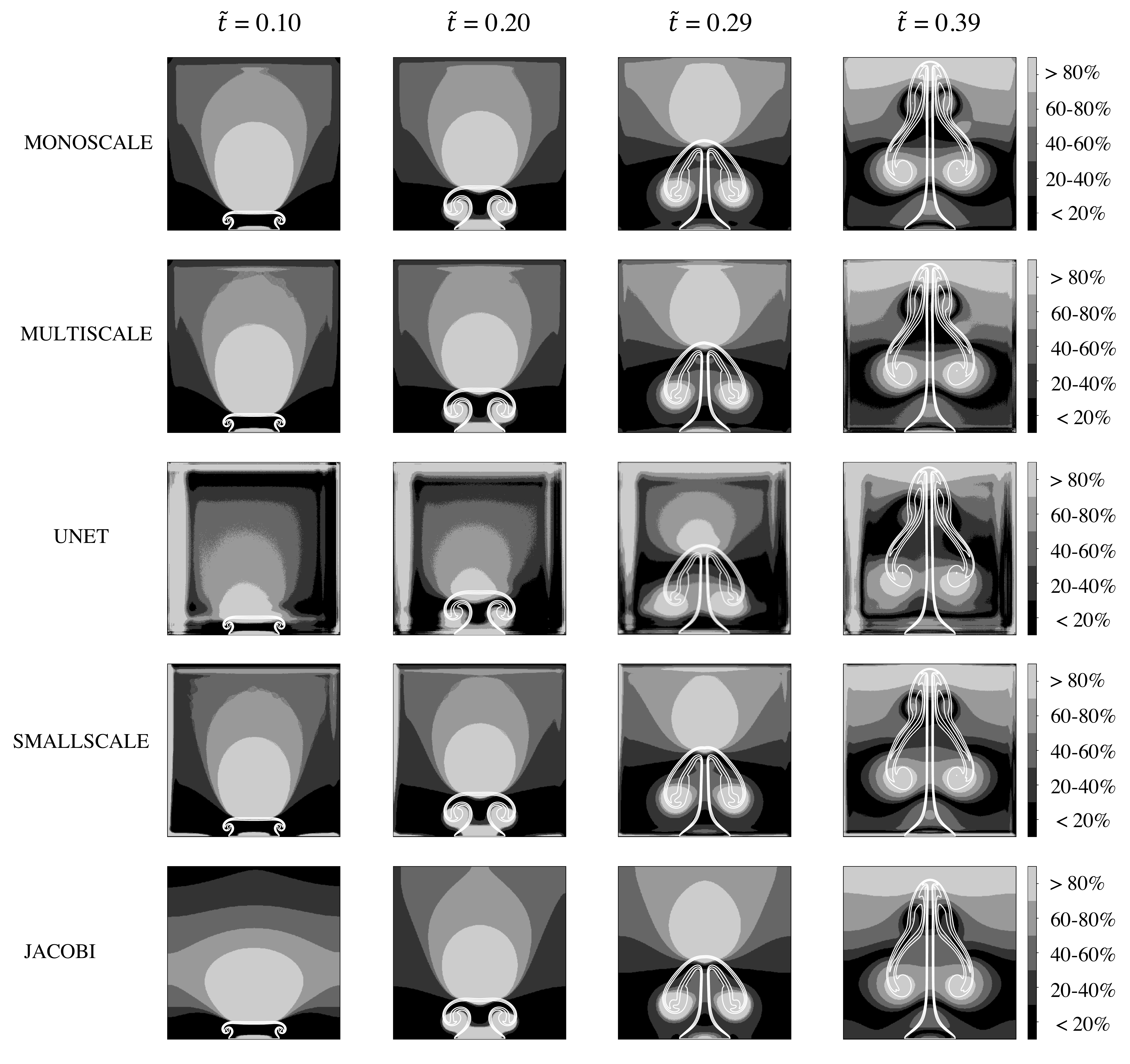}
    \caption{Divergence percentiles and density iso-contours (in white) of the four studied networks for the no cylinder case at timesteps $\tilde{t}$~=~0.10, 0.20, 0.29 and 0.39 when the networks prediction are coupled with the hybrid strategy.
    The chosen threshold is $\mathcal{E} = \mathcal{E}_{1}$, where $\mathcal{E}_{t} =  min(\mathcal{E}_{1})$ obtained by any network at a given timestep, without being coupled with the hybrid strategy.}
    \label{fig:heat_map_mean_div_no_cylinder}
\end{figure}

\begin{figure}[h!]
    \centering
    \includegraphics[width=0.99\textwidth]{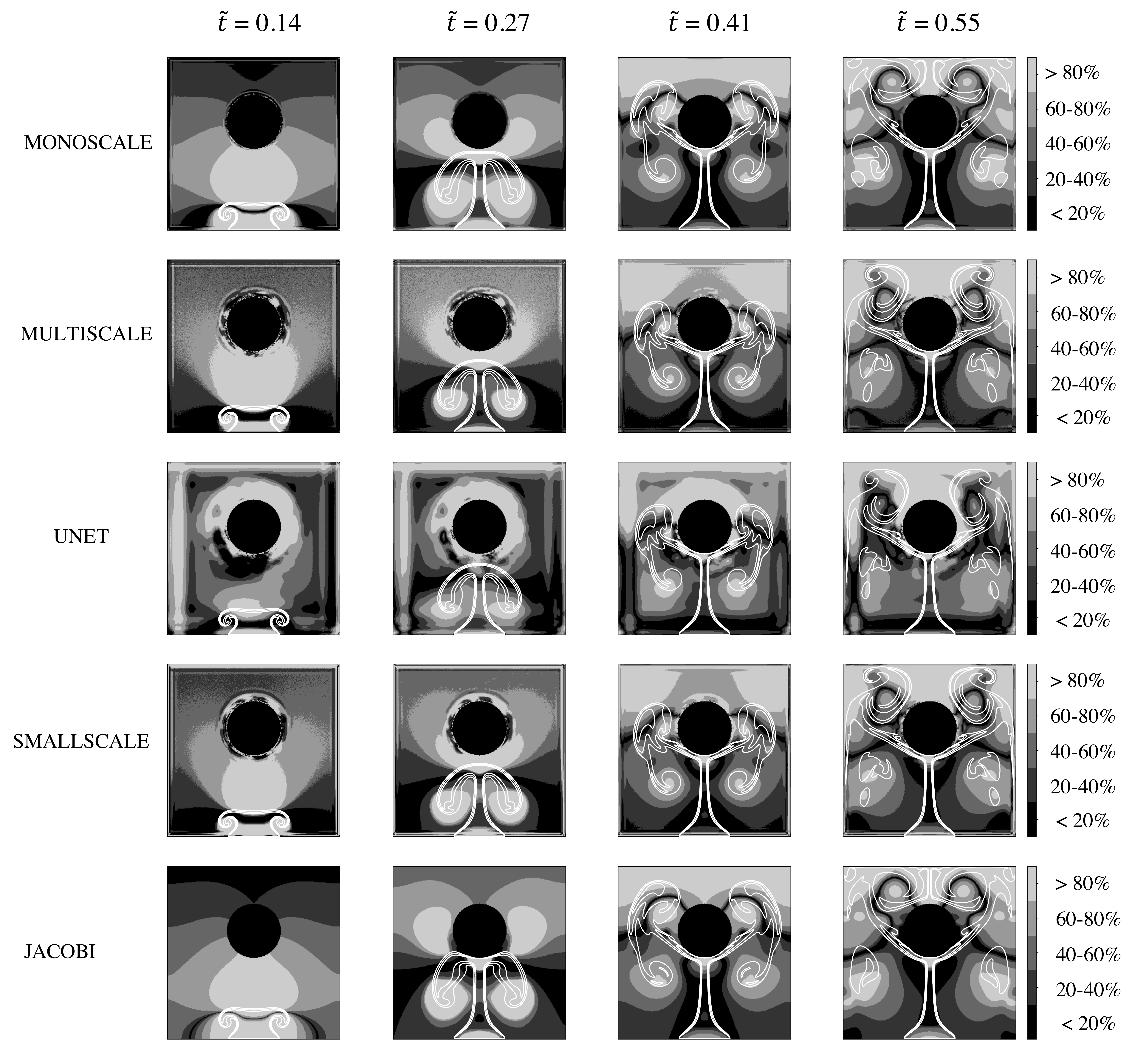}
    \caption{Divergence percentiles and density iso-contours (in white) of the four studied networks for the cylinder case at timesteps $\tilde{t}$~=~0.14, 0.27, 0.41 and 0.55 when the networks prediction are coupled with the hybrid strategy. The chosen threshold is $\mathcal{E} = \mathcal{E}_{\infty}$, where $\mathcal{E}_{t} =  min(\mathcal{E}_{\infty})$ obtained by any network at a given timestep, without being coupled with the hybrid strategy.}
    \label{fig:heat_map_max_div}
\end{figure}

\begin{figure}[h!]
    \centering
    \includegraphics[width=0.99\textwidth]{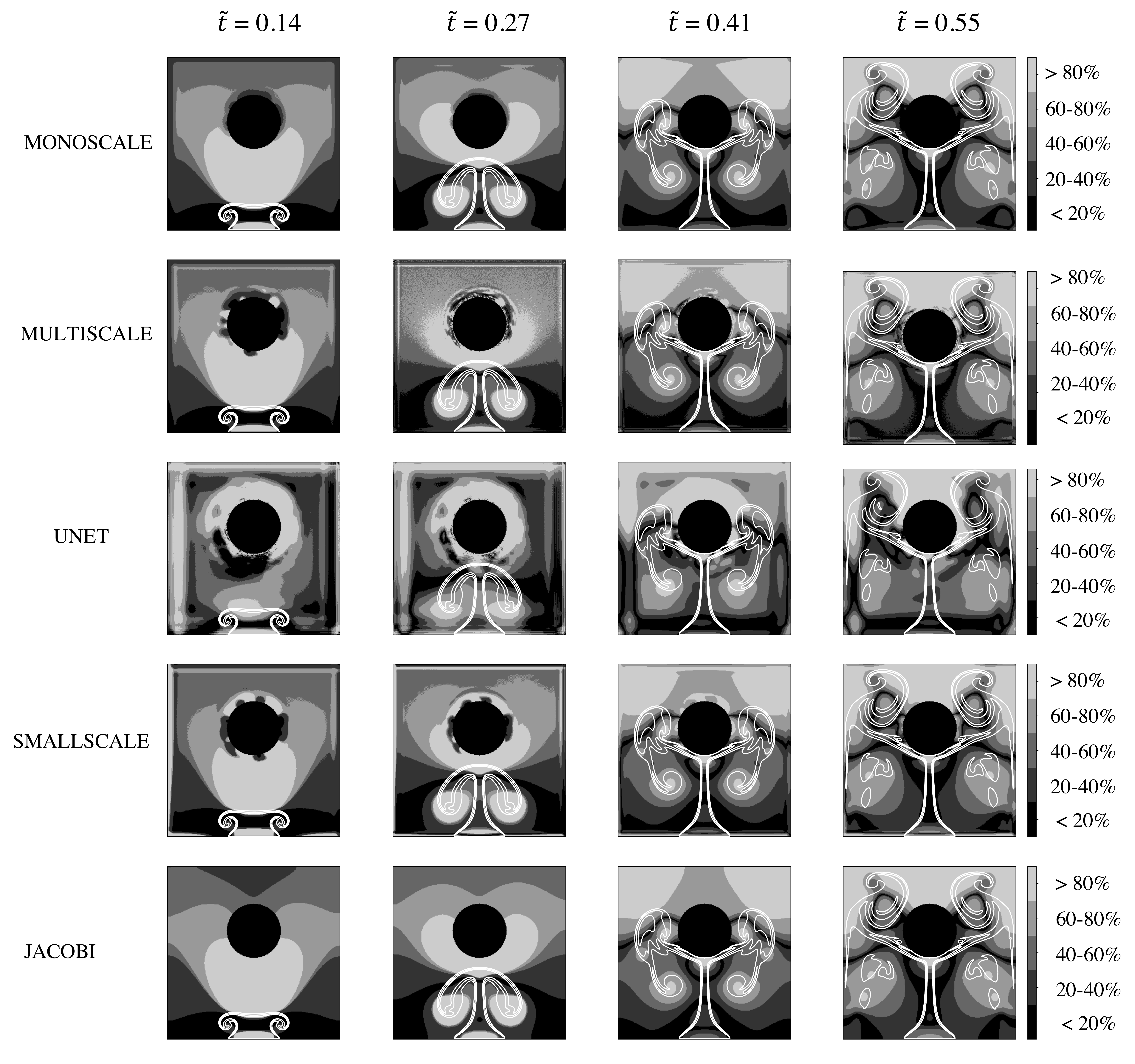}
    \caption{Divergence percentiles and density iso-contours (in white) of the four studied networks for the cylinder case at timesteps $\tilde{t}$~=~0.14, 0.27, 0.41 and 0.55 when the networks prediction are coupled with the hybrid strategy. The chosen threshold is $\mathcal{E} = \mathcal{E}_{1}$, where $\mathcal{E}_{t} =  min(\mathcal{E}_{1})$ obtained by any network at a given timestep, without being coupled with the hybrid strategy.}
    \label{fig:heat_map_mean_div}
\end{figure}

Figures~\ref{fig:heat_map_max_div_no_cylinder}-\ref{fig:heat_map_mean_div} show the divergence distribution and density iso-contours of the four studied networks, as well as the Jacobi solver with the hybrid approach.  Figures~\ref{fig:heat_map_max_div_no_cylinder}~and~\ref{fig:heat_map_mean_div_no_cylinder} correspond to the no cylinder test case with $\mathcal{E}=\mathcal{E}_{\infty}$ (Fig.~\ref{fig:heat_map_max_div_no_cylinder}) and $\mathcal{E}=\mathcal{E}_{1}$ (Fig.~\ref{fig:heat_map_mean_div_no_cylinder}). As the threshold level is set to the minimum value obtained by any networks without any Jacobi iterations, the Unet network is usually the one needing the least number of Jacobi iterations. Thus, the Unet's output closely matches the raw output, which shows the error concentration on the top and left wall, yet the global flow topology is consistent with other networks.

Figures~\ref{fig:heat_map_max_div}~and~\ref{fig:heat_map_mean_div} correspond to the cylinder test case for both thresholds $\mathcal{E}=\mathcal{E}_{\infty}$ (Fig.~\ref{fig:heat_map_max_div}) and $\mathcal{E}=\mathcal{E}_{1}$ (Fig.~\ref{fig:heat_map_mean_div}). As the less accurate network, the MonoScale network is mainly controlled by the Jacobi iterations, which results in similar flow topologies between the MonoScale and the Jacobi solver. As previously mentioned, the Unet network concentrates its error around the cylinder, as it struggles to correctly predict the pressure field around objects. This results in a quite different error distribution, even when $\mathcal{E}=\mathcal{E}_{1}$, as even if the Unet network needs some Jacobi iterations to reach the specified divergence threshold, the resulting error field is still controlled by the error concentrated around the object and walls. However, and as previously mentioned, even if the divergence distribution is different, the flow topology perfectly matches the resulting flow of the other networks.

So to sum up, it is clear that following the maximum error with the hybrid approach yields different plume behaviors, with a too fast rising for the MonoScale network and the Jacobi method, whereas the mean error provides consistent numerical simulations whatever the network architectures. For all cases and methods, several zones of high errors can be distinguished: (i) at the inlet where the lighter fluid is injected, (ii) above the plume head, and (iii) at the core of the vortices produced by the baroclinic torque, in particular at later times ($\tilde{t}=0.29$ and $0.39$) for the no cylinder test case. In that context, the Unet is producing additional error patterns, with high errors close to the boundaries, either near the cylinder walls or close to the CFD domain edges. Moreover, the error distribution in space is not symmetric, compared with all other methods. Note however that percentiles do not reflect the absolute error level, and the Unet has shown a good overall accuracy for both cases, even in the presence of the cylindrical obstacle.

\clearpage

\section{Sensitivity of the hybrid solver on the threshold value $\mathcal{E}_t$}
\label{sec:appendix_threshold}

\begin{figure}[h!]
    \centering
    \includegraphics[width =1.0\textwidth]{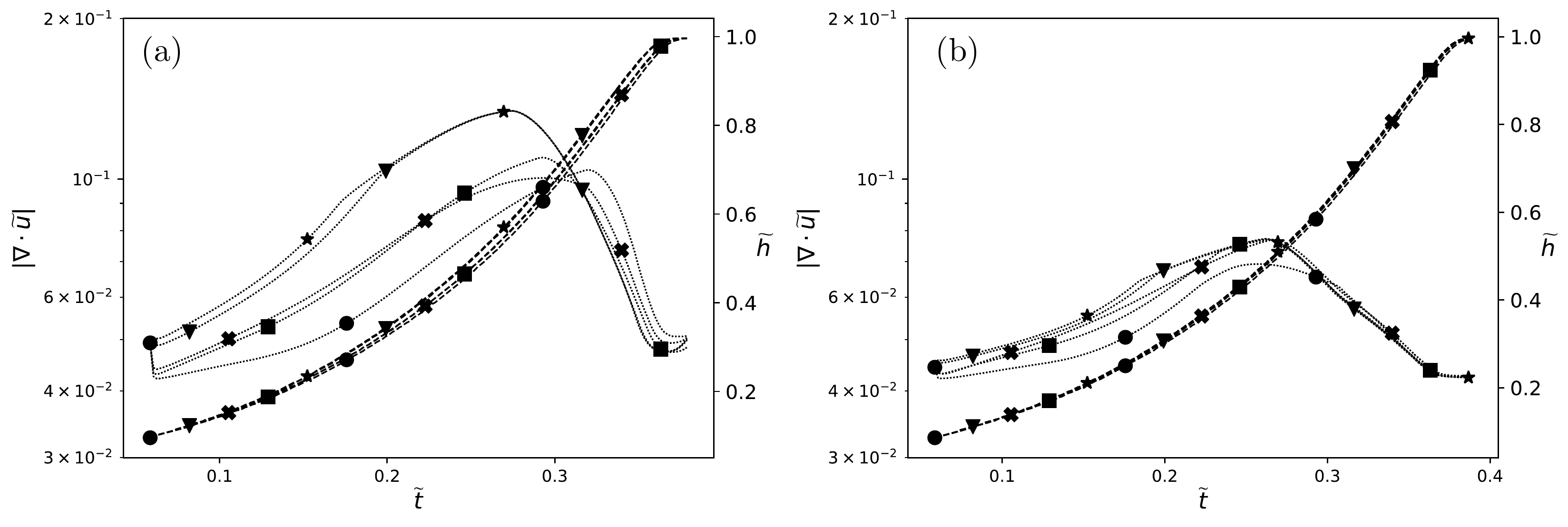}
    \caption{Plume head position $\tilde{h}_y$ (\dashed) and mean divergence of the velocity field (\dotted) for the case with $\mathcal{E}_{\infty} = 0.37$ (a) and $\mathcal{E}_{\infty} = 0.18$ (b) without an obstacle at $R_i=14.8$ obtained by several networks: $\blacktriangledown$ MonoScale, $\blacksquare$ MultiScale, $\times$ SmallScale and $\bullet$ Unet as well as the $\star$ Jacobi solver .}
    \label{fig:plume_head_hyb_thresh}
\end{figure}

%{\color{red} Je pense que tout ca va plutôt en Appendice pour le coup, ainsi que Fig. 15 et 16 ... pour gagner de la place tu dois pouvoir superposer les 2 figures 15 et 16 avec en noir les courbes de h, et en gris les distributions d'erreur}
The evolution of the head position with a lower threshold is shown in Fig.~\ref{fig:plume_head_hyb_thresh} for the plume test case without obstacle. If the maximum threshold level is set to $\mathcal{E}_t = 0.37$ (Fig.~\ref{fig:plume_head_hyb_thresh}-a), which corresponds to a divergence threshold lower than the maximum threshold that could be obtained with any network (Fig.~\ref{fig:Basic_div}), the plume does not follow a homogeneous behavior. In particular, the Jacobi method and MonoScale network produce a higher divergence level. If the threshold is further decreased by a factor $2$ ($\mathcal{E}_t = 0.18$, Fig.~\ref{fig:plume_head_hyb_thresh}-b), all methods now produce simulations with the same evolution of the plume head position. To further understand the condition which dominates the flow behavior, the mean divergence of the velocity field at each time step is displayed with dotted lines in Fig.~\ref{fig:plume_head_hyb_thresh}. It shows the evolution of $\mathcal{E}_{1}$ for both threshold values $\mathcal{E}_t = 0.37$ (a) and $\mathcal{E}_t = 0.18$ (b). As the maximum threshold is decreased, the spatially-averaged error gets similar for all networks, resulting in the same plume evolution in time. Note that discrepancies are still observed between these errors, the Unet outperforming all other networks.

%{\color{red} Dans la Figure (a) il y a eu un petit problème avec le calcul du Jacobi, qui est en process de se refaire (c'est pour ça le comportement bizarre de la tête)}

%\begin{figure}[h!]
%    \centering
%    \includegraphics[width =1.0\textwidth]{images/Hybrid/Plume/Ri_10/Threshold/Thresholds_plume_head.pdf}
%    \caption{Plume's head position $\tilde{h}_y$ (dashed lines) for the case with $\mathcal{E} = max(|\nabla \cdot {\bf u}|) = \mathcal{E}_{t}$, where $\mathcal{E}_{t} = 0.37$ (on the left) and $\mathcal{E}_{t} = 0.18$ (on the right) without an obstacle at $R_i=14.8$ obtained by several networks: $\blacktriangledown$ MonoScale, $\blacksquare$ MultiScale, $\times$ SmallScale, and $\bullet$ Unet.}
%    \label{fig:plume_head_hyb_thresh}
%\end{figure}

%\begin{figure}[h!]
%    \centering
%    \includegraphics[width = 1.0\textwidth]{images/Hybrid/Plume/Ri_10/Threshold/Thresholds_div.pdf}
%    \caption{Mean (dotted lines) divergence of the velocity field for the case for the case with $\mathcal{E} = max(|\nabla \cdot {\bf u}|) = \mathcal{E}_{t}$, where $\mathcal{E}_{t} = 0.37$ (on the left) and $\mathcal{E}_{t} = 0.18$ (on the right) without an obstacle at $R_i=14.8$ obtained by several networks: $\blacktriangledown$ MonoScale, $\blacksquare$ MultiScale, $\times$ SmallScale, and $\bullet$ Unet.}
%    \label{fig:div_mean_Hybrid}
%\end{figure}

\clearpage

\section{Evolution of the error distribution for the plume-cylinder case}
\label{sec:appendix_cylinderKDE}

\begin{figure}[h!]
    \centering
    \includegraphics[width=0.99\textwidth]{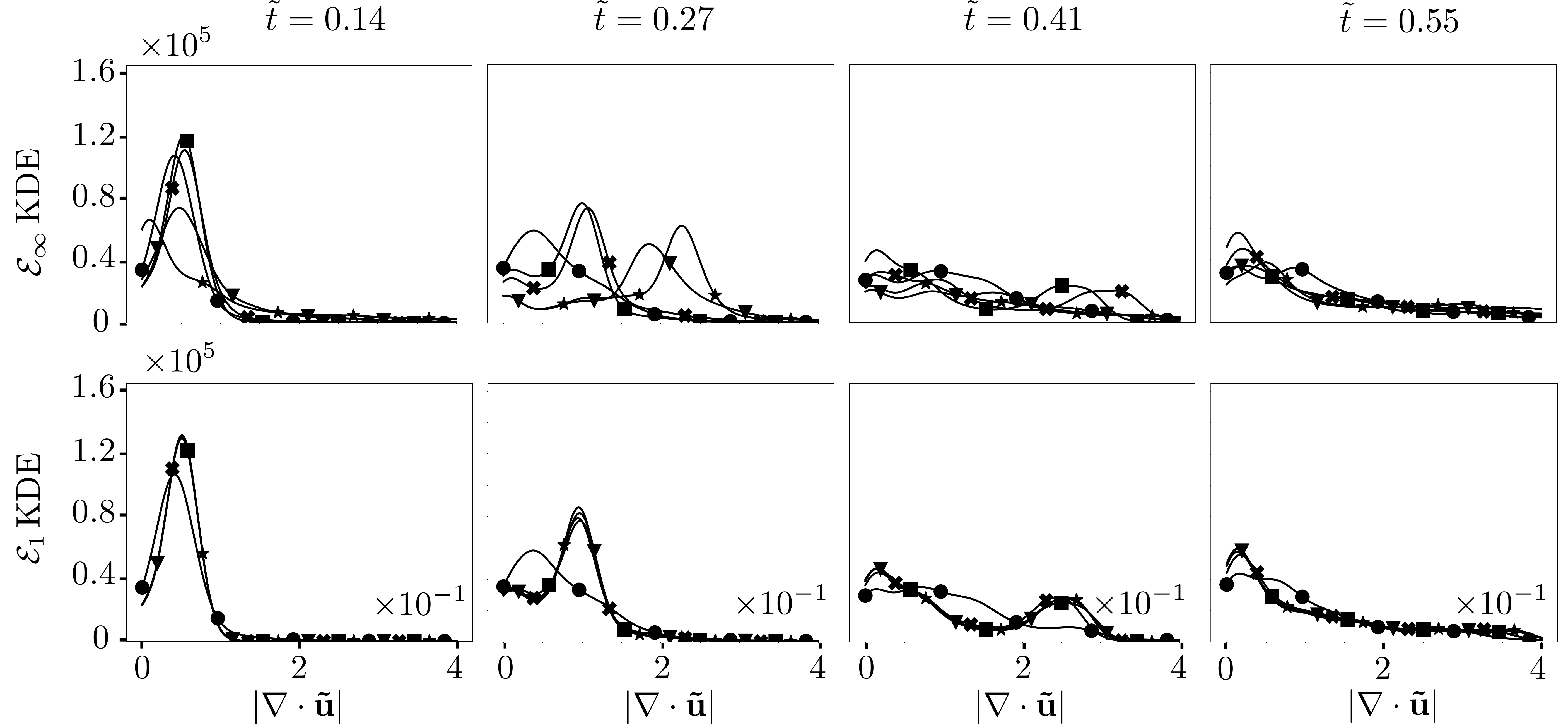}
    \caption{KDE at 4 times ($\tilde{t}$ = 0.14, 0.27, 0.41 and 0.55) of the cases where $\mathcal{E} = \mathcal{E}_{\infty}$ (top) and $\mathcal{E} = \mathcal{E}_{1}$ (bottom) of the cylinder test case, at a $R_i=14.8$ obtained by several networks: $\blacktriangledown$ MonoScale, $\blacksquare$ MultiScale, $\times$ SmallScale, and $\bullet$ Unet, as well as the $\star$ Jacobi solver.}
    \label{fig:KDE_mean_or_max_cyl}
\end{figure}

Figure.~\ref{fig:KDE_mean_or_max_cyl} shows the KDE for the plume impinging a cylinder, for the cases where the threshold $\mathcal{E}$ is equal to both $\mathcal{E}_{\infty}$ and $\mathcal{E}_{1}$ for four timesteps ($\tilde{t} = 0.14$, $0.27$, $0.41$ and $0.55$). As previously mentioned, when  $\mathcal{E}=\mathcal{E}_{\infty}$, the four networks, as well as the Jacobi solver, show a different behavior. This can be particularly appreciated at timestep $\tilde{t} = 0.27$, although all the networks follow a unimodal distribution, the MonoScale network and the Jacobi solver, previously identified as the less accurate solvers, have a divergence peak around $|\nabla \cdot {\bf u}| = 0.2$, whereas the MultiScale and SmallScale networks behave similarly with a peak around $|\nabla \cdot {\bf u}| = 0.1$, and the Unet network outperforms the rest with a peak around $|\nabla \cdot {\bf u}| = 0.05$. When $\mathcal{E}=\mathcal{E}_{1}$, the MonoScale, MultiScale, SmallScale and Jacobi solver follow an almost identical distribution, showing a divergence peak at around $|\nabla \cdot {\bf u}| = 0.1$. The Unet behavior should be highlighted, since its distribution has a wider peak, centered on around $|\nabla \cdot {\bf u}| = 0.05$. This is related to the threshold level, which follows the lowest value of the mean divergence of all the studied networks. At timestep $\tilde{t} = 0.27$, this value corresponds to the Unet network, so to achieve this threshold level, the other networks rely on Jacobi iterations which homogenize the divergence distribution, whilst the Unet's KDE corresponds to the network's output. Despite this difference, the plume flow is not affected, as macroscopically the flow structures are identical for all the cases when $\mathcal{E}= \mathcal{E}_{1}$.

\end{document}